\documentclass[11pt]{article}

\usepackage{graphicx}
\usepackage{amssymb}
\usepackage{amsmath}
\usepackage{natbib}
\usepackage[pdftex, pdfusetitle, plainpages=false,
				letterpaper, bookmarks, bookmarksnumbered,
				colorlinks, linkcolor=black, citecolor=black,
	         filecolor=black, urlcolor=black]
				{hyperref}

\usepackage[all,arc]{xy}
\usepackage{enumerate}
\usepackage{mathrsfs}
\usepackage{setspace}
\usepackage{multirow}
\usepackage{algorithm,algpseudocode}
\algnewcommand{\Inputs}[1]{%
  \State \textbf{Inputs:}
  \Statex \hspace*{\algorithmicindent}\parbox[t]{.8\linewidth}{\raggedright #1}
}
\algnewcommand{\Initialize}[1]{%
  \State \textbf{Initialize:}
  \Statex \hspace*{\algorithmicindent}\parbox[t]{.8\linewidth}{\raggedright #1}
}

\usepackage{JASA_manu}

\newtheorem{thm}{Theorem}[section]

\newtheorem{lem}[thm]{Lemma}

\begin{document}

\title{Modular Bayes screening for high-dimensional predictors}

\author{
Yuhan Chen  \\
Department of Statistical Science \\ Duke University, Durham, NC 27708  \\
email: \texttt{yuhan.chen@duke.edu}
  \and
David Dunson  \\
Department of Statistical Science \\ Duke University, Durham, NC 27708  \\
email: \texttt{dunson@duke.edu}
}
\maketitle

\begin{center}
\textbf{Abstract}
\end{center}
With the routine collection of massive-dimensional predictors in many application areas, screening methods that rapidly identify a small subset of promising predictors have become commonplace.  We propose a new MOdular Bayes Screening (MOBS) approach, which involves several novel characteristics that can potentially lead to improved performance.  MOBS first applies a Bayesian mixture model to the marginal distribution of the response, obtaining posterior samples of mixture weights, cluster-specific parameters, and cluster allocations for each subject.  Hypothesis tests are then introduced, corresponding to whether or not to include a given predictor, with posterior probabilities for each hypothesis available analytically conditionally on unknowns sampled in the first stage and tuning parameters controlling borrowing of information across tests.  By marginalizing over the first stage posterior samples, we avoid under-estimation of uncertainty typical of two-stage methods. We greatly simplify the model specification and reduce computational complexity by using {\em modularization}. We provide basic theoretical support for this approach, and illustrate excellent performance relative to competitors in simulation studies and the ability to capture complex shifts beyond simple differences in means. The method is illustrated with applications to genomics by using a very high-dimensional cis-eQTL dataset with roughly 38 million SNPs.
\vspace*{.1in}


\section{Introduction}

In modern scientific research, it has become routine to collect massive dimensional data for each study subject, leading to a massive $p$ small to moderate $n$ statistical problem. In most scientific studies collecting high-dimensional data, the over-arching interest is not in developing a black-box model for prediction; instead the focus is on variable selection. In particular, the scientists would like to select the subset of features out of a high-dimensional set of $p$ candidates $x_{ij}$, which are predictive of a response $y_i$, with $i=1,\ldots,n$ indexing the subject under study. There is a vast recent literature on high-dimensional feature/variable selection, with the overwhelming focus on one of two strategies. The first is to use a penalized optimization method, such as Lasso (\citealt{Tib}), SCAD (\citealt{FanLi}), Dantzig selector (\citealt{Candes}) or other variants, to obtain a sparse estimate of the regression coefficient vector. In general, such an approach has adequate performance only when the feature matrix $X$ satisfies a number of stringent conditions, which seldom hold in scientific applications even approximately. When these assumptions are violated, performance can degrade badly.

This fact has lead to an emphasis on the second strategy, which is to first apply an independent screening algorithm using measures of association between $x_{ij}$ and $y_i$ separately for $j=1,\ldots,p$, and then select features based on thresholding the corresponding p-values, test statistics or association estimates to maintain a superset containing most of the important features along with a number of false positives. This strategy was proposed by \cite{FanLv} using marginal correlation for sure independence screening (SIS). \cite{FanSong} extended the idea to generalized linear models and \citet{FanFengSong} considered nonparametric additive models. \cite{Zhu} introduced model-free screening via sure independent ranking and screening (SIRS). Other model-free procedures include the distance correlation screening (DCS) in \citet{Li} and the fused Kolmogorov filter in \cite{Mai} and \cite{Mai2}.

The focus of this article is on improving upon high-dimensional screening methodology through using Bayesian nonparametric and hierarchical models to avoid parametric assumptions on the data generating model, while borrowing information across the many tests being conducted. A potentially key disadvantage of most existing screening methods is the lack of any borrowing of information across the different tests. A novel idea we propose, which should have impact beyond the screening case, is to place a nonparametric Bayes model on the {\em marginal} distribution of the response, with inclusion tests for each predictor then defined conditionally on unknowns in the marginal distribution. This approach has the dual advantages of avoiding parametric assumptions on the distribution of the response, while also automatically borrowing information across the screening tests through their shared conditioning on a common set of unknowns. In addition to this novel type of borrowing, we additionally place a more convention Bayesian hierarchical structure on the probability of variable inclusion, related to \cite{Scott}. A key advantage of the proposed formulation is that we can characterize uncertainty in posterior computation for the marginal response distribution via integrating conditional posterior probabilities of variable inclusion across first stage posterior samples. This is computationally scalable and avoids the under-estimation of uncertainty typical of two-stage procedures..

One caveat about our proposed approach is that it is not a coherent fully Bayes probability model in that we ignore certain dependencies for simplicity, robustness and computational tractability. In particular, to define a fully Bayes model for the marginal distribution of the response $y$ in settings involving predictors $x_j = (x_{1j},\ldots,x_{nj})'$, we would require a model for the conditional distribution of the response $y$ given all the predictors $x$ along with a model for the joint distribution of the predictors $x$. In very large $p$ settings, computational and complexity considerations force a focus on relatively simple settings involving strong constraints (e.g., Gaussian, linear, highly sparse, etc). In most settings, such constraints are inconsistent with available prior information, and results are therefore highly questionable. Instead, we take the more mild approach of ignoring information on $x$ in defining the marginal distribution of $y$. This is a type of {\em modularization} (\citealt{Liu}). The idea of modularization is that Bayesian models can be defined in {\em modules}, with posterior computation in certain modules not taking into account the model structure and data in other modules. In our case, the model for the marginal density $f(y)$ is one module, and we do not attempt to take into account information about the predictors in fitting this module. We suspect that closely related ideas may be a game changer in making (approximately) Bayesian approaches practical in modern massive dimensional data settings.

The existing literature on nonparametric Bayes variable selection is sparse. There are several methods for two group comparisons (\citealt{Dun, Ma, Ho}), but such methods are computationally intensive for each group, and would need to be applied separately $p$ times. There are also several approaches that are designed for variable selection in Bayesian nonparametric models for conditional response distributions, $f(y|x)$. For example, \cite{Chung} proposed a stochastic search variable selection algorithm under a probit stick-breaking mixture model for $f(y|x)$. \citet{Kess} and \cite{Yang} used tensor factorizations to characterize the conditional distribution in order to capture complex interactions between the predictors. \citet{Jiang} introduced a nonparametric test using a nonparametric Bayesian slice inversed regression based on modeling the conditional distribution of a covariate given the discretized response. Their method is less computationally intensive than previous attempts, but is restricted to univariate $y$.  Finally, \citet{Filippi} tested for pairwise dependence between two variables using Dirichlet process mixtures, improving efficiency by running MCMC for each marginal in parallel.

In Section 2 we propose the general MOdular Bayes Screening (MOBS) framework, provide basic theoretical support, and sketch a general approach to computation. In Section 3 we focus on the case in which the response is continuous and univariate, motivating a location-scale mixture of Gaussians for the marginal distribution. Section 4 contains a simulation study assessing operating characteristics relative to a variety of high-dimensional screening methods. Section 5 applies the screening method to a massive-dimensional cis-eQTL dataset with approximately 38 million SNPs. Section 6 contains a discussion. Technical details are included in an Appendix.

\section{Modular Bayes screening framework}
\subsection{General framework}
MOdular Bayes Screening (MOBS) uses a two stage set-up and starts with a mixture model for the marginal density $f(y)$ of the response. For subjects $i=1,\ldots,n$, let $y_i$ denote the response, which can be either univariate or multivariate. Let $x_i = (x_{i1},\ldots,x_{ip})'$ denote a high-dimensional vector of categorical predictors for subject $i$, with $x_{ij} \in \{0,1,\ldots,d_j-1\}$ having $d_j$ levels. Suppose that marginally we have $y_i \sim f$ independently for $i=1,\ldots,n$; we refer to $f$ as the {\em baseline} density. Treating this baseline density as unknown, we apply a finite over-fitted mixture model (\citealt{Rousseau}):
\begin{eqnarray}
f(y) = \sum_{h=1}^k \omega_h \mathcal{K}( y| \theta_h), \label{eq:mixture}
\end{eqnarray}
$$(\omega_1,\cdots,\omega_k)' \sim \mbox{Dir}\bigg( \frac{\alpha}{k},\ldots,\frac{\alpha}{k}\bigg),\quad \theta_h \sim P_0,$$ where $\mathcal{K}$ is the kernel density, $\omega = (\omega_1,\cdots,\omega_k)'$ are the weights, $\theta = (\theta_1,\cdots,\theta_k)'$ the kernel parameters, $P_0$ is the prior distribution for each $\theta_h$ and $k$ is a conservative upper bound on the number of mixture components needed to produce an accurate approximation of the unknown density. Let $c = (c_1,\ldots,c_n)'$ denote a latent variable vector with $c_i \in \{1,\ldots,k\}$ being the cluster membership of $y_i$ and $\mbox{Pr}(c_i=h) = \omega_h$. As $k$ increases, (\ref{eq:mixture}) converges weakly to a Dirichlet process mixture model. This represents a common approximation to the Dirichlet Process mixture model (\citealt{Ishwaran}) facilitating practical implementation.

To generalize the baseline density to allow dependence on the $j$th predictor $x_j$, we let
\begin{eqnarray}
f(y|x_j) = \sum_{h=1}^k \omega_{jh}(x_j) \mathcal{K}(y| \theta_{jh}(x_j)), \label{eq:mixxj}
\end{eqnarray}
where $\omega_j(l) = (\omega_{j1}(l),\cdots,\omega_{jk}(l))'$ and $\theta_j(l) = (\theta_{j1}(l),\cdots,\theta_{jk}(l))'$, $l \in \{0,1,\ldots,d_j-1\}$, represent the weights and component parameters for subjects with $x_{ij}=l$. Expression (\ref{eq:mixxj}) modifies the baseline density (\ref{eq:mixture}) to allow the weights and kernel parameters to vary with $x_j$. Let
\begin{eqnarray}
H_{0j}: \omega_j(0)=\ldots=\omega_j(d-1) = \omega, \quad \theta_j(0)=\ldots=\theta_j(d-1) = \theta, \label{eq:Hj2}
\end{eqnarray}
$$H_{1j}: \mbox{Not all } \omega_j(l) \mbox{ are the same and/or not all } \theta_j(l) \mbox{ are the same}$$
correspond to discarding and including the $j$th predictor, respectively. The alternative hypothesis $H_{1j}$ is composed of three cases: $H_{1j1}$, where $\omega$ varies but $\theta$ remains the same, $H_{1j2}$, where $\theta$ varies but $\omega$ is the same and $H_{1j3}$, where both vary. By testing both the weights and the cluster parameters, MOBS is able to capture deeper and more complex shifts in distribution than traditional methods that focus primarily on the mean. Clearly, $H_{1j} = H_{1j1} \cup H_{1j2} \cup H_{1j3}$. Let $\kappa = (\kappa_{0}, \kappa_{11}, \kappa_{12}, \kappa_{13})'$ be the prior probabilities of the four subhypotheses, where $\kappa_{0}$ is the prior probability of $H_{0j}$ and $\kappa_{11}, \kappa_{12}, \kappa_{13}$ are the prior probabilities of the alternative hypotheses $H_{1j1}, H_{1j2}, H_{1j3}$ respectively.
We suppose that the weight and component parameter vectors specific to each level of each predictor are distributed randomly about their baseline values according to the hierarchical model:
\begin{eqnarray}
\omega_j(l) \sim \mbox{Dir}( \tau_\omega \omega ), \quad \theta_{jh}(l) \sim P_1, \label{eq:priorbase}
\end{eqnarray}
where $P_1$ is a distribution centered on $\theta$ with scale parameter $\tau_\theta$ and $\tau_\omega$ is the Dirichlet precision controlling variability about the mean $\omega$. Using a Dirichlet prior for $\omega_j(l)$ maintains conjugacy with the Multinomial distribution of the component memberships $c$ and facilitates marginalizing out $\omega_j(l)$ when calculating the likelihood. For the same reason, $P_1$ should ideally be conjugate to the kernel density $\mathcal{K}$. The precisions $\tau=(\tau_{\omega},\tau_{\theta})$ control how much $f(y|x_j)$ changes with predictor $x_j$ for each $j$ such that $H_{1j}$ holds.  The hierarchical structure favors borrowing of information in learning $f(y|x_j)$ across different predictors and levels of $x_j$.


\subsection{Algorithmic details}

Using (\ref{eq:Hj2}) allows direct calculation of an analytic form for $\mbox{Pr}(H_{0j}|y,x_j,\Lambda,\Phi)$, where $\Phi$ denotes $\omega,c,\theta$, the baseline weights and component parameters,  and $\Lambda$ denotes $\kappa,\tau_\omega,\tau_\theta$, the hyperparameters controlling borrowing of information and multiplicity adjustment in hypothesis testing. The key idea in our proposed MOBS approach is to run MCMC for posterior computation only for the baseline nonparametric model for the marginal response density $f(y)$. We do not know the exact kernel memberships, weights or parameters, and there does not exist a tractable, analytic form for $\mbox{Pr}(H_{0j}|y,x_j,\Lambda)$, so MOBS instead takes the Monte Carlo integration of $\mbox{Pr}(H_{0j}|y,x_j,\Lambda,\Phi)$ over the samples of $\Phi$ generated from the posterior distribution under the baseline model.

Let $\mbox{L}(c, y|H_{1j}, x_j, \omega, \theta, \Lambda)$ and $\mbox{L}(c, y|H_{0j}, x_j, \omega, \theta, \Lambda)$ be the likelihood conditional on the baseline MCMC output and hyperparameters controlling borrowing of information and multiplicity adjustment in hypothesis testing for $H_{1j}$ and $H_{0j}$, respectively. Then, we have
$$\mbox{L}(c,y|H_{1j}, x_j, \omega, \theta, \Lambda) = \sum_{t=1}^3 \frac{\kappa_{1t}}{1-\kappa_0} \mbox{L}(c,y|H_{1jt}, x_j, \omega, \theta, \Lambda),$$ where $\mbox{L}(c,y|H_{1jt}, x_j, \omega, \theta, \Lambda)$ is the conditional likelihood for $H_{1jt}$ for $t = 1,2,3$. The posterior probabilities of $H_{0j}$ and $H_{1jt}$ conditional on the baseline weights and parameters are then, respectively,
\begin{eqnarray}
\mbox{Pr}(H_{0j}|\kappa,y,x_j,\Phi)&=&\frac{1}{1+\frac{\kappa_{11}}{\kappa_{0}}\mbox{BF}_{11}(j)+\frac{\kappa_{12}}{\kappa_{0}}\mbox{BF}_{12}(j)+\frac{\kappa_{13}}{\kappa_{0}}\mbox{BF}_{13}(j)},\label{eq:postj}\\
\mbox{Pr}(H_{1jt}|\kappa,y,x_j,\Phi)&=&\frac{\kappa_{1t}}{\kappa_{0}}\mbox{BF}_{1t}(j)\mbox{Pr}(H_{0j}|\kappa,y,x_j,\Phi),\label{eq:post1jt}
\end{eqnarray}
where for $t=1,2,3$, $$\mbox{BF}_{1t}(j) = \frac{\mbox{L}(c,y|H_{1jt}, x_j, \omega, \theta, \Lambda)}{\mbox{L}(c,y|H_{0j}, x_j, \omega, \theta, \Lambda)}$$ is the Bayes factor in favor of $H_{1jt}$ over $H_{0j}$ conditional on the hyperparameters and the baseline weights and parameters.

We calculate the conditional likelihoods given the baseline unknowns and hyperparameters as follows. First, under $H_{0j}$, we have the simple form:
\begin{eqnarray*}
\mbox{L}(c,y|H_{0j}, x_j, \omega, \theta, \Lambda)&=&\mbox{L}(c|H_{0j}, x_j, \omega, \theta, \Lambda) \mbox{L}(y|H_{0j}, x_j, \Phi, \Lambda)\\
&=& \prod_{h=1}^k  \omega_h^{n_{jh}}\prod_{i: c_i=h} \mathcal{K}( y_i| \theta_h),
\end{eqnarray*}
where $n_j(l)=(n_{j1}(l),\ldots,n_{jk}(l))'$, with $n_{jh}(l)$  the number of subjects having $x_{ij}=l$ and belonging to cluster $h$, and  $n_j=(n_{j1},\ldots,n_{jk})'=\sum_{l=0}^{d_j-1} n_j(l)$  the total number of subjects allocated to each component. For the three alternative subhypotheses
\begin{eqnarray*}
\mbox{L}(c,y|H_{1j1}, x_j, \omega, \theta, \Lambda)&=&\mbox{L}(c|H_{1j1}, x_j, \omega, \theta, \Lambda) \mbox{L}(y|H_{1j1}, x_j, \Phi, \Lambda)\\
&=& \left[\prod_{l=0}^{d_j-1}\int \left\{\prod_{h=1}^k\omega_{jh}(l)^{n_{jh}(l)}\right\}\mbox{Dir}(\omega_{j}(l)|\tau_\omega \omega) d\omega_{j}(l)\right]  \prod_{h=1}^k\prod_{i: c_i=h} \mathcal{K}( y_i| \theta_h),\\
\mbox{L}(c,y|H_{1j2}, x_j, \omega, \theta, \Lambda)&=&\mbox{L}(c|H_{1j2}, x_j, \omega, \theta, \Lambda) \mbox{L}(y|H_{1j2}, x_j, \Phi, \Lambda)\\
&=&\prod_{l=0}^{d_j-1}\left\{\prod_{h=1}^k\omega_{h}^{n_{jh}}\right\}\int\prod_{h=1}^k P_1(\theta_{jh}(l))\prod_{i:c_i=h} \mbox{L}(y_i|\Phi,\Lambda, \theta_j(l))d\theta_{jh}(l),\\
\mbox{L}(c,y|H_{1j3}, x_j, \omega, \theta, \Lambda)&=&\mbox{L}(c|H_{1j3}, x_j, \omega, \theta, \Lambda)
\mbox{L}(y|H_{1j3}, x_j, \Phi, \Lambda)
\end{eqnarray*}
\begin{eqnarray*}
&=&\left[\prod_{l=0}^{d_j-1}\int \left\{\prod_{h=1}^k\omega_{jh}(l)^{n_{jh}(l)}\right\}\mbox{Dir}(\omega_{j}(l)|\tau_\omega \omega) d\omega_{j}(l)\right]\\
 && \prod_{l=0}^{d_j-1}\int\prod_{h=1}^k P_1(\theta_{jh}(l))\prod_{i:c_i=h} \mbox{L}(y_i|\Phi,\Lambda, \theta_j(l))d\theta_{jh}(l).\\
\end{eqnarray*}

These results imply after some calculations that
\begin{eqnarray*}
\mbox{BF}_{11}(j)&=&\frac{\prod_{l=0}^{d_j-1}\int \left\{\prod_{h=1}^k\omega_{jh}(l)^{n_{jh}(l)}\right\}\mbox{Dir}(\omega_{j}(l)|\tau_\omega \omega) d\omega_{j}(l) }{\omega_1^{n_1}\omega_2^{n_2}\ldots\omega_k^{n_k}},\\
\mbox{BF}_{12}(j)&=&\frac{ \prod_{l=0}^{d_j-1}\int\prod_{h=1}^k P_1(\theta_{jh}(l))\prod_{i:c_i=h} \mbox{L}(y_i|\Phi,\Lambda, \theta_j(l))d\theta_{jh}(l)}{\prod_{h=1}^k  \prod_{i: c_i=h} \mathcal{K}( y_i| \theta_h)},\\
\mbox{BF}_{13}(j)&=&\frac{ \prod_{l=0}^{d_j-1}\int\prod_{h=1}^k P_1(\theta_{jh}(l))\prod_{i:c_i=h} \mbox{L}(y_i|\Phi,\Lambda, \theta_j(l))d\theta_{jh}(l)}{\prod_{h=1}^k  \prod_{i: c_i=h} \mathcal{K}( y_i| \theta_h)}\\
&& \frac{\prod_{l=0}^{d_j-1}\int \left\{\prod_{h=1}^k\omega_{jh}(l)^{n_{jh}(l)}\right\}\mbox{Dir}(\omega_{j}(l)|\tau_\omega \omega) d\omega_{j}(l)}{\omega_1^{n_1}\omega_2^{n_2}\ldots\omega_k^{n_k}}\\
 &=& \mbox{BF}_{11}(j)\mbox{BF}_{12}(j).
\end{eqnarray*}
Here, $\mbox{BF}_{11}(j)$ provides a weight of evidence of changes in mixture weights with $x_j$, while $\mbox{BF}_{12}(j)$ provides the same for different kernels. $\mbox{BF}_{13}(j)$ provides evidence for both weights and kernels being different and interestingly ends up being the product of $\mbox{BF}_{11}(j)$ and $\mbox{BF}_{12}(j)$.

\begin{algorithm}
  \caption{Modular Bayes Screening}
  \begin{algorithmic}[1]
    \Inputs{data $X, y$}
    \Initialize{$\kappa_0, \kappa_{11}, \kappa_{12}, \kappa_{13}, \tau_\mu, \tau_\theta$}
    \State Obtain samples for $\Phi$ from the posterior distribution under baseline mixture model using MCMC or an alternative algorithm.
    \Repeat
      \State Estimate $\mbox{Pr}(H_{0j}|y,x_j,\Lambda)$ and $\mbox{Pr}(H_{1jt}|y,x_j,\Lambda)$ for $t=1,2,3$ by averaging $\mbox{Pr}(H_{0j}| y,x_j,\Lambda,\Phi)$ and $\mbox{Pr}(H_{1jt}| y,x_j,\Lambda,\Phi)$, respectively, over the baseline $\Phi$ samples for each $j$.
      \State Set $\kappa_0, \kappa_{11}, \kappa_{12}, \kappa_{13}$ using the average of the $\mbox{Pr}(H_{0j}|\kappa,y,x_j,\Lambda)$, $\mbox{Pr}(H_{1j1}|y,x_j,\Lambda)$, $\mbox{Pr}(H_{1j2}|y,x_j,\Lambda)$, and $\mbox{Pr}(H_{1j3}|y,x_j,\Lambda)$ estimated above, respectively.
    \Until{convergence}
  \end{algorithmic}
\end{algorithm}

We initialize $\kappa_{0} = 0.5$, $\kappa_{11} = \kappa_{12} = \kappa_{13} = 0.5/3$. Estimating $\kappa$ involves iteratively taking the average of the posterior probabilities for each subhypothesis, which is effectively an empirical Bayes procedure. Our proposed MOBS approach leverages on the simplicity of posterior computation for $f(y)$ and the simple analytic forms shown above via Algorithm 1.

\section{Univariate continuous response}

\subsection{Computation}
While the proposed method allows for complex, arbitrary responses, this section provides an example under a simple univariate, continuous $y$ setting. The baseline density $f(y)$ is represented as a mixture of univariate location-scale Gaussian kernels:
\begin{eqnarray}
f(y) &=& \sum_{h=1}^k \omega_h \mbox{N}( y| \mu_h, \sigma^2_h), \label{eq:mixuni}
\end{eqnarray}
where $\{\mu_h, \sigma^2_h\} = \theta_h$ are the mean and variance parameters of the component $h$ with prior $P_0 = \mbox{N}(\mu_h|\mu_0, q\sigma_h^2)\mbox{IGa}(\sigma_h^2|a, b)$. Details on the Gibbs sampler for the base distribution can be found in the Appendix. The conditional density $f(y|x_j)$ can similarly be represented as
\begin{eqnarray}
f(y|x_j) &=& \sum_{h=1}^k \omega_{jh}(x_j) \mbox{N}(y| \mu_{jh}(x_j), \sigma^2_{jh}(x_j)),  \label{eq:mixunixj}
\end{eqnarray}
where $\{\mu_{jh}(x_j), \sigma^2_{jh}(x_j)\} = \theta_{jh}(x_j)$ with prior $P_1 = \mbox{N}(\mu_{jh}(l)|\mu_h, \sigma^2_{jh}(l)/\tau_\mu)\mbox{IGa}(\sigma^2_{jh}(l)|a_h, b_h),$ with $a_h=\frac{\tau_\sigma}{\sigma_h^4}$, $b_h=\frac{\tau_\sigma}{\sigma_h^2}$ and the scale parameters $\tau_\theta = \{\tau_\mu, \tau_\sigma\}$.

To derive $\mbox{BF}_{11}(j)$, note that for each $j$ and $l$,
\begin{eqnarray*} & &\int \left\{\prod_{h=1}^k \omega_{jh}(l)^{n_{jh}(l)}\right\}\mbox{Dir}(\omega_{j}(l)|\tau_\omega \omega) d\omega_{j}(l)= \frac{\beta(n_j{(l)}+\tau_\omega\omega)}{\beta(\tau_\omega\omega)},
\end{eqnarray*}
where $\beta$ is the multivariate beta function.  Hence
\begin{equation}
\mbox{BF}_{11}(j)=\frac{1}{\omega_1^{n_1}\omega_2^{n_2}\ldots\omega_k^{n_k}}\prod_{l=0}^{d_j-1} \frac{\beta(n_j{(l)}+\tau_\omega\omega)}{\beta(\tau_\omega\omega)}
\label{eq:bf1}.\end{equation}

Similarly to find $\mbox{BF}_{12}(j)$ for each $j$, first note that
$$\prod_{h=1}^k  \prod_{i: c_i=h} \mathcal{K}( y_i| \theta_h)=\frac{1}{(2\pi)^{\frac{n}{2}}}\prod_{h=1}^k\frac{1}{\sigma_h^{n_h}}e^{-\frac{\sum_{i:c_i=h} (y_i-\mu_h)^2}{2\sigma^2_h}}$$
and
\begin{eqnarray*}
 &&\prod_{l=0}^{d_j-1}\int\prod_{h=1}^k P_1(\theta_{jh}(l))\prod_{i:c_i=h} \mbox{L}(y_i|\Phi,\Lambda, \theta_j(l))d\theta_{jh}(l)\\
& = & \prod_{l=0}^{d_j-1} \int \mbox{L}(y|c,\omega,\mu_j(l),\sigma_j^2(l)) p(\sigma^2_{jh}(l)|a_h,b_h)p(\mu_{jh}(l)|\mu_h, \tau_\mu, \sigma^2_{jh}(l)) d\mu_j(l) d\sigma_j^2(l)\\
&= & \frac{1}{(2\pi)^{\frac{n}{2}}} \prod_{l=0}^{d_j-1} \prod_{h=1}^K\frac{\Gamma(a_{jh}(l))}{\Gamma(a_h)}\frac{b_h^{a_h}}{b_{jh}(l)^{a_{jh}(l)}}\frac{\sqrt{\tau_\mu}}{\sqrt{\tau_\mu+n_{jh}(l)}},
\end{eqnarray*}
where \begin{eqnarray*}
 &&a_{jh}(l)=a_h+\frac{n_{jh}(l)}{2}, \quad \bar{y}_{jh}(l)=\frac{1}{n_{jh}(l)}\sum_{i:c_i=h, x_{ij}=l} y_i,\\
 &&b_{jh}(l)=b_h+\frac{n_{jh}(l)\tau_\mu}{2(\tau_\mu+n_{jh}(l))}\{\mu_h-\bar{y}_{jh}(l)\}^2 +\sum_{i:c_i=h, x_{ij}=l} \frac{\{y_i-\bar{y}_{jh}(l)\}^2}{2}.
 \end{eqnarray*}
It follows that
\begin{eqnarray}
&&\mbox{BF}_{12}(j)=\left(\prod_{l=0}^{d_j-1} \prod_{h=1}^k\frac{\Gamma(a_{jh}(l))}{\Gamma(a_h)}\frac{b_h^{a_h}}{b_{jh}(l)^{a_{jh}(l)}}\frac{\sqrt{\tau_\mu}}{\sqrt{\tau_\mu+n_{jh}(l)}}\right) \left(\prod_{h=1}^k \sigma_h^{n_h}e^{\frac{\sum_{c_i=h} (y_i-\mu_h)^2}{2\sigma^2_h}}\right)\label{eq:bf2}.\end{eqnarray}

Running MOBS requires first generating samples of $c,\omega,\mu$ and $\sigma^2$ from the posterior under the baseline mixture model. One then iteratively alternates between estimation of $\mbox{Pr}(H_{0j}|y,x_j,\Lambda)$ using (\ref{eq:bf1}) and (\ref{eq:bf2}) over the samples of the baseline $c,\omega,\mu$ and $\sigma^2$ and estimation of $\kappa$ as the average of the estimated posterior probabilities. The strength of MOBS lies in the ease of computing $\mbox{BF}_{11}(j)$ and $\mbox{BF}_{12}(j)$ in a trivially parallelizable manner. The overall algorithm has computational complexity $O(np)$. The approach can be trivially modified to accommodate multivariate and discrete settings; in such cases, expression (\ref{eq:bf2}) will take a different form.

\subsection{Hyperparameters selection}

An important question when running MOBS is the selection of hyperparameters, especially that of the precision parameters $\tau = \{\tau_\mu, \tau_\sigma, \tau_\omega\}$. The various $\tau$ can be interpreted as the precisions controlling the distance between the density of $y$ under the null hypothesis and that of $y|x_j=l$ under the alternative hypothesis. Naturally, higher values of $\tau$ would suggest a smaller distance, as $y|x_j=l$ converges to $y$ when $\tau_\mu, \tau_\sigma, \tau_\omega \rightarrow \infty$.

We aim to tune $\tau$ such that the prior signal-to-noise ratio is within a reasonable range somewhere between $0.05$ and $0.1$. For any fixed $j=1,2,...,p$ and $l=0, 1,...,d_j-1$, let $\tilde{\Phi}=\{\mu, \sigma, \omega, \mu_{j}(l), \sigma_{j}(l), \omega_{j}(l)\}$. Given $\tilde{\Phi}$, the densities of $\mu, y$ and $y|x_j=l$ are
\begin{eqnarray*}
f_{\mu|\tilde{\Phi}}(t)= \sum_{h=1}^k \omega_h\mbox{N}(t|\mu_0, q\sigma^2_h),\\
f_{y|\tilde{\Phi}}(t)= \sum_{h=1}^k \omega_h \mbox{N}(t|\mu_h, \sigma^2_h),\\
f_{y|x_j=l, \tilde{\Phi}}(t)= \sum_{h=1}^k \omega_{jh}(l) \mbox{N}(t|\mu_{jh}(l), \sigma^2_{jh}(l)).
\end{eqnarray*}

Let
$$\Delta_0=E\left[\|f_{y|\tilde{\Phi}}-f_{\mu|\tilde{\Phi}}\|_2^2\right],$$
$$\Delta_1=E\left[\|f_{y|\tilde{\Phi}}-f_{y|x_j=l,\tilde{\Phi}}\|_2^2\right]$$
be the average square of $L_2$-norm distances between the densities of $\mu|\tilde{\Phi}$ and $y|\tilde{\Phi} $ and between those of $y|\tilde{\Phi}$ and $y|x_j=l, \tilde{\Phi}$ respectively. Since $\int \mbox{N}(t|\mu_1, \sigma_1^2)\mbox{N}(t|\mu_2,\sigma_2^2)dt=N\left(\mu_1-\mu_2|0,\sigma_1^2+\sigma_2^2\right)$, we know
\begin{eqnarray*}
\|f_{y|\tilde{\Phi}}-f_{\mu|\tilde{\Phi}}\|_2^2&=& \sum_{h=1}^k\sum_{t=1}^k \omega_h\omega_t \left[\mbox{N}(0|0, q(\sigma_h^2+\sigma_t^2))+\mbox{N}(\mu_h-\mu_t|0,\sigma_h^2+\sigma_t^2)\right.\\
&&\left.-2 \mbox{N}(\mu_h-\mu_0|0, \sigma_h^2+q\sigma_t^2)\right],
\end{eqnarray*}
\begin{eqnarray*}
\|f_{y|\tilde{\Phi}}-f_{y|x_j=l,\tilde{\Phi}}\|_2^2&=& \sum_{h=1}^k\sum_{t=1}^k \omega_h\omega_t \mbox{N}(\mu_h-\mu_t|0, \sigma_h^2+\sigma_t^2)\\
&&+\sum_{h=1}^k\sum_{t=1}^k \omega_{jh}(l)\omega_{jt}(l) \mbox{N}(\mu_{jh}(l)-\mu_{jt}(l)|0,\sigma^2_{jh}(l)+\sigma^2_{jt}(l))\\
&&-2\sum_{h=1}^k\sum_{t=1}^k \omega_h\omega_{jt}(l)\mbox{N}(\mu_h-\mu_{jt}(l)|0, \sigma_h^2+\sigma_{jt}(l)^2).
\end{eqnarray*}

While $\Delta_0$ and $\Delta_1$ are both intractable, we can estimate their values by averaging $\|f_{y|\tilde{\Phi}}-f_{\mu|\tilde{\Phi}}\|_2^2$ and $\|f_{y|\tilde{\Phi}}-f_{y|x_j=l,\tilde{\Phi}}\|_2^2$ over samples of $\tilde{\Phi}$. Empirically, we find that for a given $k$, setting $\mu_0 = 0, \alpha=k, a=2, b=0.02, q=50, \tau_\mu=50, \tau_\sigma=50, \tau_\omega=k^{3/2}+8(k-1)$ provides a reasonable default specification that is relatively stable and has a signal-to-noise ratio $\Delta_1/\Delta_0$ between 0.05 and 0.1.

\subsection{Theoretical properties}

We initially study asymptotic properties of MOBS treating parameters $\Phi$ as known, and then consider sure screening consistency under general settings. We first investigate properties of $\mbox{Pr}(H_{0j}|y,x_j,\Lambda,\Phi)$. The posterior odds conditioned on $\Phi$ and $\Lambda$ are
\begin{eqnarray*}
\frac{\mbox{Pr}(H_{1j}|y,x_j,\Lambda,\Phi)}{\mbox{Pr}(H_{0j}|y,x_j,\Lambda,\Phi)}&=& \frac{\kappa_{11}}{\kappa_{0}}\mbox{BF}_{11}(j)+\frac{\kappa_{12}}{\kappa_{0}}\mbox{BF}_{12}(j)+\frac{\kappa_{13}}{\kappa_{0}}\mbox{BF}_{12}(j) \mbox{BF}_{11}(j).
\end{eqnarray*}
Set $d_j=2$. For $l=0, 1$, $h=1,2,\ldots,k $ and $j=1,2,\ldots,p$, let $$n_j^l=\sum_{h=1}^k n_{jh}(l), \quad\quad  \lambda_{0}=\frac{n_{j}^0}{n_{j}^0+n_{j}^1},\quad \quad p_{jhl}=\frac{n_{jh}(l)}{n_j^l}.$$
Therefore, $n_j^0=n \lambda_0$ and $n_j^1=n(1-\lambda_0)$ where $n=n_j^0+n_j^1$. Let $\Omega_{j}(l)$ and $\Theta_j(l)=(M_j(l), \Sigma^2_{j}(l))$ be the true values of $\omega_{j}(l)$ and $\theta_j(l)=(\mu_j(l), \sigma^2_j(l))$, respectively, for $j=1,2,\ldots,p, l=0,1$. Then under $H_{0j}$, $\Omega_{j}(l)=\omega$, $M_j(l)=\mu$ and $\Sigma^2_j(l)=\sigma^2$.

\begin{thm} Let $n_j^0,n_j^1\rightarrow \infty$, with $\frac{n_j^0}{n_j^1} = \frac{\lambda_0}{1-\lambda_0}$ for some fixed $\lambda_0$. Treat $\Phi$ as known.

(1) When $H_{0j}$ is true, we have $\mbox{BF}_{11}(j) \rightarrow 0$ with rate $n^{-(k-1)}$ and
$\mbox{BF}_{12}(j) \rightarrow 0$ with rate $n^{-2k}$. These two estimates imply that $\mbox{Pr}(H_{1j}|y,x_j,\Lambda,\Phi)\rightarrow 0$ with rate $n^{-(k-1)}$.

(2) Under the condition $\Omega_j(0)\neq \Omega_j(1)$, we have $\log\mbox{BF}_{11}(j)\rightarrow \infty$ with rate $An$ and under the condition $\Theta_j(0)\neq \Theta_j(1)$, we have $\log\mbox{BF}_{12}(j)\rightarrow \infty$ with rate $Bn$ where
\begin{eqnarray*}
A&=&\sum_{h=1} \left[ \lambda_0\Omega_{jh}(0)\log\left\{\frac{\Omega_{jh}(0)}{ \omega_h}\right\}+\lambda_1\Omega_{jh}(1)\log\left\{\frac{\Omega_{jh}(1)}{ \omega_h}\right\}\right]\\
B&=&\frac{1}{2}\sum_{h=1}^k\sum_{l=0}^1 \frac{\lambda_l\Omega_{jh}(l)\{M_{jh}(l)-\mu_h\}^2}{\sigma_h^2}+\frac{1}{2}\sum_{h=1}^k\sum_{l=0}^1\lambda_l\Omega_{jh}(l)\left[\frac{\Sigma^2_{jh}(l)}{\sigma_h^2}-1+\log\left\{\frac{\sigma_h^2}{\Sigma^2_{jh}(l)}\right\}\right].
 \end{eqnarray*}
 Therefore under $H_{1j}$, $\mbox{Pr}(H_{0j}|y,x_j,\Lambda,\Phi)\rightarrow 0$ with rate $e^{-n \max(A, B)}$.
\end{thm}

The above theorem shows rates of convergence of hypothesis probabilities and Bayes factors for the different hypotheses in the event that the true weights, parameters and cluster allocations are known. However, in practice MOBS uses  posterior samples for the baseline parameters to account for uncertainty. Now consider the situation where the parameters are unknown and the model may be misspecified. In particular, we extend our above results using misspecification techniques given in \cite{Klein}  and used in \cite{Lock}. Let $\mathbb{F}$ be the set of all convex combinations of Gaussian distributions $\{f_h(\cdot|\theta_h)\}$ where $\theta_h=\{\mu_h, \sigma^2_h\}$ and let $P$ define a prior on $\mathbb{F}$. Let $f_0$ be the true distribution and $f^*$ the closest convex combination in $\mathbb{F}$ to $f_0$ under Kullback-Leibler divergence.  We define $B(\epsilon,f^{*}|f_0)$ to be a neighborhood of the density $f^*$ under the measure induced by the density $f_0$: $$B(\epsilon,f^{*}|f_0) =\left \{f \in \mathbb{F}: -\int f_0 \log \frac{f}{f^{*}} \leq \epsilon^2, \int f_0 \log\left(\frac{f}{f^{*}}\right)^2 \leq \epsilon^2\right\}$$ and define $d(f_1,f_2)$ to be the weighted Hellinger distance $$d^2(f_1,f_2) = \frac{1}{2} \int (f_1^{1/2}-f_2^{1/2})^2 \frac{f_0}{f^{*}}.$$

\begin{lem} Let $y_1, \ldots , y_n$ be independent with density $f_0$. Assume $f^*= \mbox{argmin}_{f \in \mathbb{F}} KL(f_0||f)$ exists and $P_n(B(\epsilon,f^{*}|f_0)) > 0$ for all $\epsilon > 0$ where $P_n$ is the posterior measure given prior $P$. Let $\omega^{*} = (\omega_1^{*},\ldots,\omega_k^{*})'$ be the component weights and $\theta^{*} = (\theta_1^{*},\ldots,\theta_k^{*})'$ be the kernel parameters corresponding to $f^{*}$ and let $\omega$ and $\theta$ be the true weights and kernel parameters. Assume $\omega^{*}$ and $\theta^{*}$ are unique in that $\sum \omega_h f_h(\cdot|\theta_h) = \sum \omega_h^{*} f_h(\cdot|\theta_h^{*}) = f^{*}$ only if $\omega = \omega^{*}, \theta = \theta^{*}$. Then for any fixed $\epsilon > 0$,
$$P_n((\omega,\theta) \in (\mathbb{S}^{k-1},\mathbb{R}^{k}): ||(\omega,\theta)-(\omega^{*},\theta^{*})|| \geq \epsilon | y) \rightarrow 0.$$
\end{lem}

\begin{thm} For $l = \{0,1\}$, let $y_i|x_{ij}=l$ denote the subset of $y_i$ with size $n_j^l$ such that $x_{ij}=l$. Assume $y_i|x_{ij}=0$ are independent with density $f^{(0)}$ and $y_i|x_{ij}=1$ are independent with density $f^{(1)}$. Let
$$f^{*(0)} = argmin_{f\in\mathbb{F}} KL(f^{(0)}||f),\quad f^{*(1)} = argmin_{f\in\mathbb{F}} KL(f^{(1)}||f).$$ Assume the uniqueness condition for the previous theorem holds for $f^{*(0)}$ and $f^{*(1)}$. If $f^{(0)} = f^{(1)}$, $\mbox{Pr}(H_{0j}|y,x_j,\Lambda) \rightarrow 1$ as $n_j^0,n_j^1\rightarrow \infty$ and if $f^{*(0)} \neq f^{*(1)}$, $\mbox{Pr}(H_{0j}|y,x_j,\Lambda) \rightarrow 0$ as $n_j^0,n_j^1\rightarrow \infty$.
\end{thm}
This result suggests that our posterior probability is consistent under $H_{0j}$ and holds under weak conditions under $H_{1j}$ for each $j=1,\ldots,p$. However, consistency can fail under $H_{1j}$ when $f^{(0)}$ and $f^{(1)}$ have the same closest point $f^*$ under K-L divergence in $\mathbb{F}$. In practice, $f^{(0)}$ and $f^{(1)}$ would need to be extremely close to have the same closest point, so this is a very mild condition.

Next, we use the previous theorem to establish sure screening consistency for MOBS. Following standard screening conventions (\citealt{Li}), let $F(y|X)$ denote the conditional distribution function of $y$ given the predictors $X$. Define $$D = \{j: F(y|X) \mbox{ depends on } x_j \mbox{ for some } y\}$$ to be the true set of relevant predictors. A good screening method can identify a small subset $S$ such that $D \subset S$.  Existing literature have focused predominantly on frequentist methods that select $S$ based on rankings or thresholding using some test statistic such as marginal correlation in SIS or distance correlation in DCS. MOBS instead relies on thresholding of the estimated posterior null probabilities $\hat{\pi}_j = \mbox{Pr}(H_{0j}|y,x_j,\Lambda)$ for $j=1,\ldots, p$. In particular, let $$\hat{D} = \{j: \hat{\pi}_j \le \hat{\pi}_{(d_n)} \},$$ where $\hat{\pi}_{(1)},\ldots,\hat{\pi}_{(p)}$ are the order values and $d_n$ is an integer. Let $\gamma_j=1$ if the $j$th predictor is marginally related to $y$ and $\gamma_j=0$ otherwise. We impose the following two conditions:

(C1) All jointly important predictors $j \in D$ are also marginally important (i.e. $\gamma_j = 1$) and $d_n \geq |D|$.

(C2) All jointly important predictors $j \in D$ satisfy the constraint that $f^{(0)}_j$ and $f^{(1)}_j$ are not the closest in K-L divergence to the same $f^*_j \in \mathbb{F}$.

Condition 1 is similar to conditions found in the screening literature (\citealt{FanLv}), restricting metrics of marginal importance for important predictors to be non-zero. Condition 2 requires Theorem 3.3 to be satisfied, and is mild as noted above.

\begin{thm} Under conditions (C1) and (C2), when $n, p \rightarrow \infty$,
$$\mbox{Pr}(D \subset \hat{D}) \rightarrow 1.$$

\end{thm}

\section{Simulation}
\subsection{Screening accuracy}
In this section, we assess the performance of MOBS against five existing screening methods using simulated datasets. The first competitor is  sure independence screening or SIS (\citealt{FanLv}), and we use the \texttt{SIS} \texttt{R} package. We also compare against three frequentist model-free methods, sure independent ranking and screening or SIRS (\citealt{Zhu}), using code found at
\emph{http://users.stat.umn.edu/~wangx346/research/example1b.txt}, the distance correlation screening or DCS (\citealt{Li}), using the \texttt{energy} \texttt{R} package and the fused Kolmogorov filter (\citealt{Mai2}) using code provided by the authors. Finally, we compare to the JYL Bayesian nonparametric test (\citealt{Jiang}) with code found at \emph{http://www.people.fas.harvard.edu/~junliu/BF/bfslice.html}.

We simulate 100 replicates for all methods under six different models. Under these settings, we test $n=200$ observations and $p=2000$ total predictors and generate $x_{ij} \in \{0,1\}$ and $y$. First, we consider the case where $X$ is independent and generate the features with $x_{ij} \sim \mbox{Bern}(0.5)$, $i=1,\ldots,n$, $j=1,\ldots,p$ for all 2000 predictors. However, in biological applications (e.g. involving SNPs), $X$ can often exhibit moderate correlations in blocks of predictors. To mimic such settings, we randomly select 600 predictors and draw them in the following fashion with correlation $\rho = 0.5$. We first draw $z=(z_{ij})_{n \times p}$ for each $z_{ij}\sim N(0,1)$ for $i=1,\ldots,n$ and $j=1,2,\ldots,p$. Next, construct $b=(b_{ij})_{n \times p}$ by replacing the first $\rho \cdot n$ rows of $z$ such that for $1\leq j\leq p$ for some constant $a$,
$$b_{ij}=\left\{
  \begin{array}{ll}
    z_{i1}+az_{ij}, &  1\leq i\leq \rho\cdot n \\
    z_{ij}, & \rho\cdot n<j\leq n
  \end{array}
\right.$$
To convert this continuous data to multinomial, for each predictor, we can simply assign them by quantiles. The remaining 1900 predictors are generated from independent Gaussians without correlation.

$\mathit{Model \,1\, (Linear \,regression,\, uncorrelated):}$ $y = 1 + 2x_1 + x_2 - 2x_3 + x_4 - 2x_5 + \epsilon$, where $\epsilon \sim \mbox{N}(0,1)$ is independent of $X$ and $X$ is uncorrelated.

$\mathit{Model \,2 \,(Linear \,regression, \,correlated):}$ $y$ is generated the same as with \textit{Model 1} but with a block of predictors with correlation described above.

$\mathit{Model\, 3 \,(Single \,index, \,uncorrelated):}$ $y = (1 + 2x_1 + x_2 - 2x_3 + x_4 - 2x_5)^2 + \epsilon$, where $\epsilon \sim \mbox{N}(0,1)$ is independent of $X$ and $X$ is uncorrelated.

$\mathit{Model\, 4\, (Single \,index, \,correlated):}$ $y$ is generated the same as with \textit{Model 3} but with a block of predictors with correlation described above.

$\mathit{Model \, 5\, (Mixture ,\, uncorrelated):}$ $y$ is generated using 6 predictors at positions chosen at random from $\{1,\ldots,p\}$. Let $S$ be the true vector of jointly important indices. The resulting marginal density is then sampled from $64$ mixtures of Gaussian distributions with mean $\mu \sim \mbox{Unif}(-1, 1)$ and standard deviation $\sigma \sim \mbox{Unif}(0, 1/8)$ each corresponding to different permutations of $S$. If $x_{iS} = (x_{iS1}, \ldots , x_{iS6})$, we subsequently draw $y_i \sim N(\mu_{x_{iS1}, \ldots , x_{iS6}}, \sigma^2_{x_{iS1},\ldots, x_{iS6}})$. $X$ is generated to be uncorrelated.

$\mathit{Model\, 6\, (Mixture ,\, correlated):}$ $y$ is generated the same as with \textit{Model 5} but with a block of predictors with correlation described above.

In order to implement MOBS, we use the computational strategy described in Section 2 and Section 3 and set the deafult hyperparameters $\mu_0=0, q=50, a=2, b=0.02, \alpha=k, \tau_\omega=k^{3/2}+8(k-1),\tau_\mu=\tau_\sigma=50$, with $k=3$ for the linear case since we expect the response to be mostly unimodal and $k=7$ for the more multimodal single index and mixture models. We run the Markov chain of the baseline model 6000 times and perform Monte Carlo integration using the last 500 samples. If potential issues involving label-switching for the samples of $\theta$ arise, we use methods found in \cite{Steph} and \cite{Papa}.

\begin{figure}[h]
\centering
\begin{minipage}{.9\textwidth}
  \includegraphics[width=5.8in]{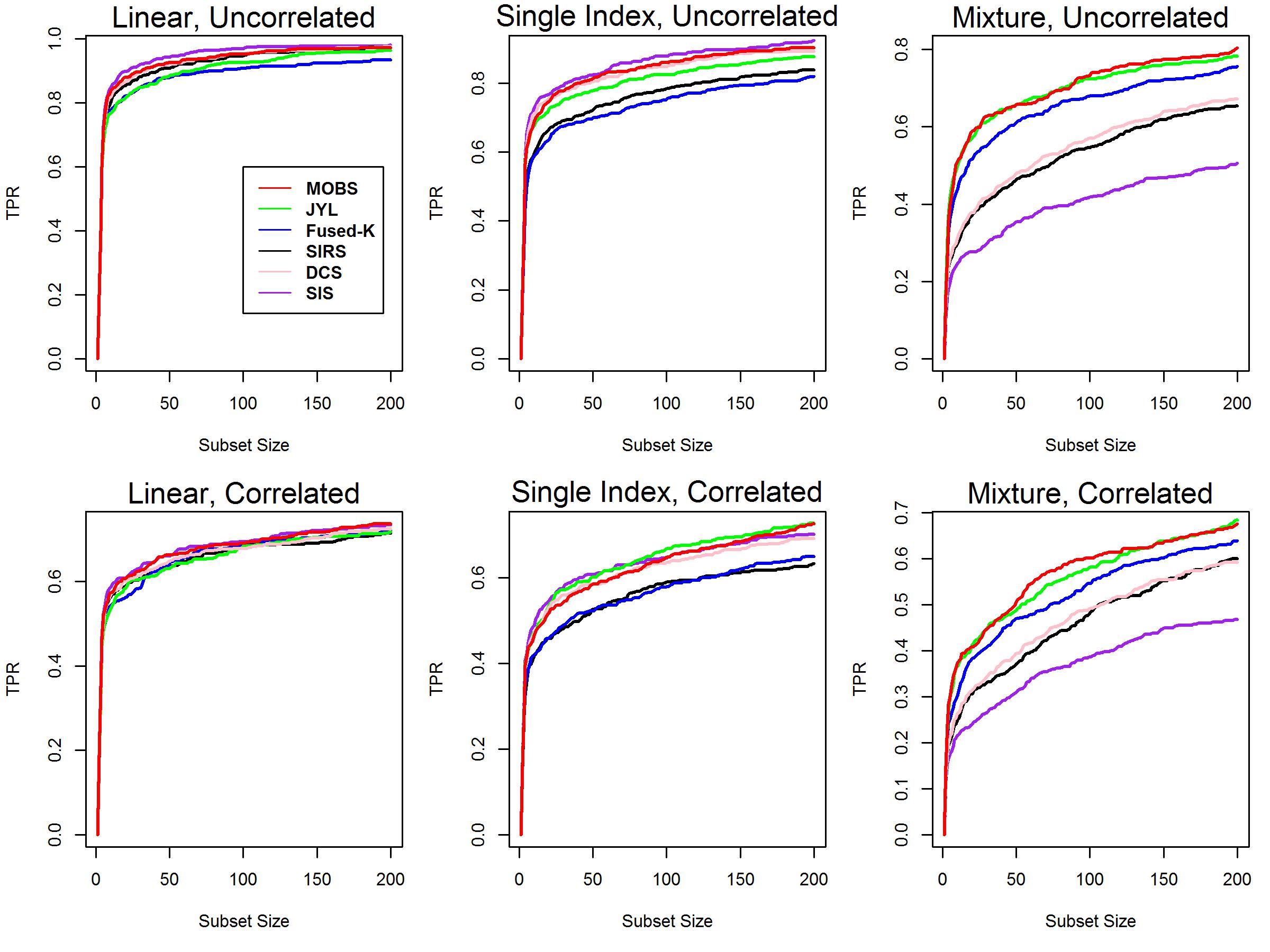}
  \caption{ROC curves for all six methods under linear, single index and mixture settings, both correlated and uncorrelated.}
\end{minipage}%
\end{figure}

Figure 1 summarizes the ROC curves for the six different methods by comparing the true positive rate against the false positive rate. In each of the simulations, MOBS exhibits the best or one of the best performance in comparison to the SIS, SIRS, DCS, the fused Kolmogorov filter and the JYL approaches. In the mixture setting, the nonparametric methods dominate as expected. For the linear regression case and single index case, the other two nonparametric methods, the fused Kolmogorov filter and JYL both demonstrate weakness in these settings, but MOBS still performs strongly and is comparable to SIS. In a multimodal mixture setting, both $\omega$ and $\theta$ play large roles in detecting change, whereas in a unimodal setting, $\omega$ is less relevant and MOBS reduces to essentially a $t$-test.

\subsection{Computational efficiency}
Understanding the computational efficiency of variable screening is important as the various methods should be scalable to huge data sets. We measure the total time taken to run each of the six methods under two different settings using data generated from an uncorrelated linear regression model setting in \textit{Model 1} but with varying $n$ and $p$. Our first example fixes the number of subjects at 250 and then alter the number of predictors. Next, we fix the number of predictors at 2500 and instead change the number of subjects.

\begin{figure}
\centering
  \includegraphics[width=5.2in]{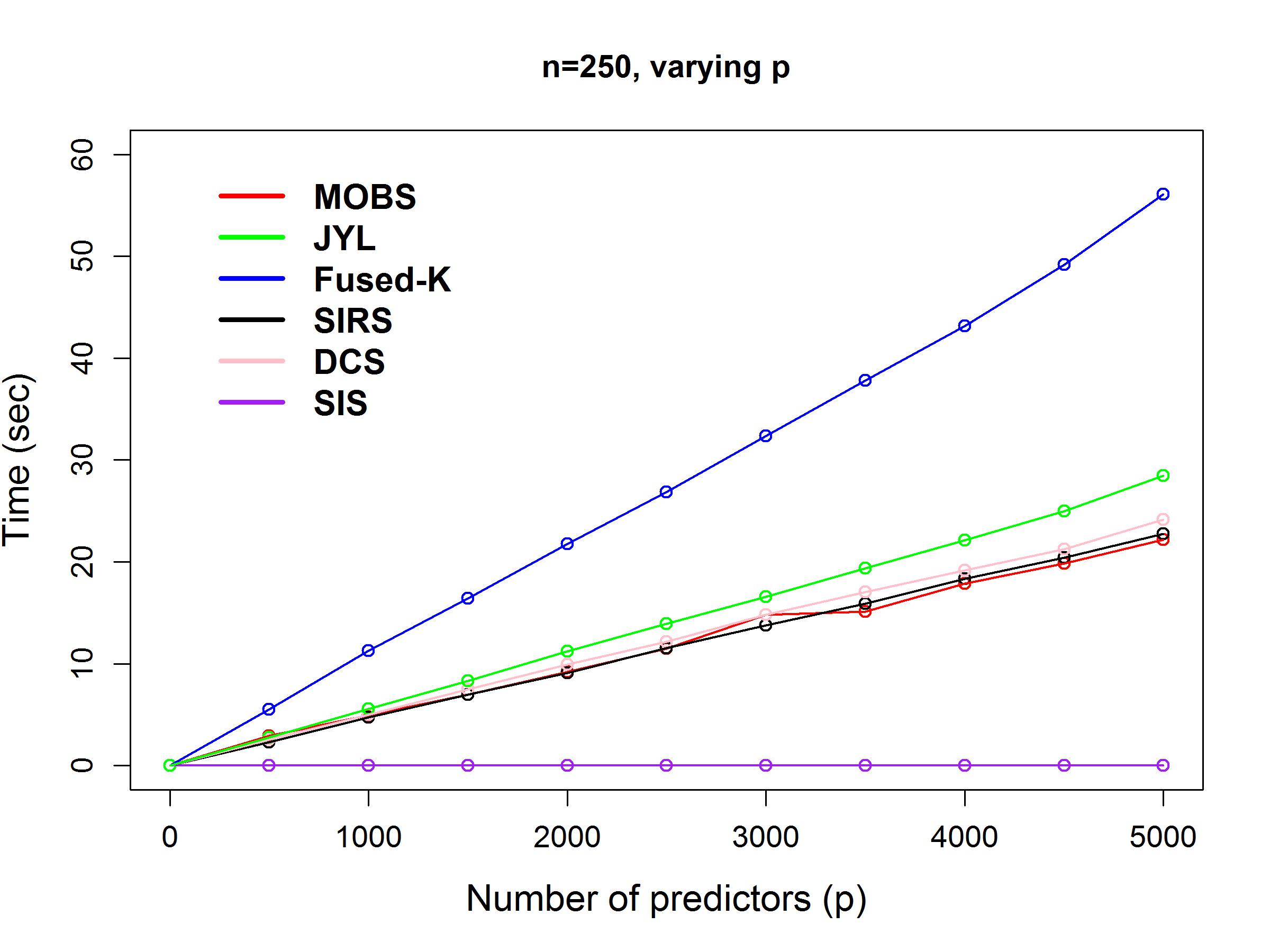}
  \includegraphics[width=5.2in]{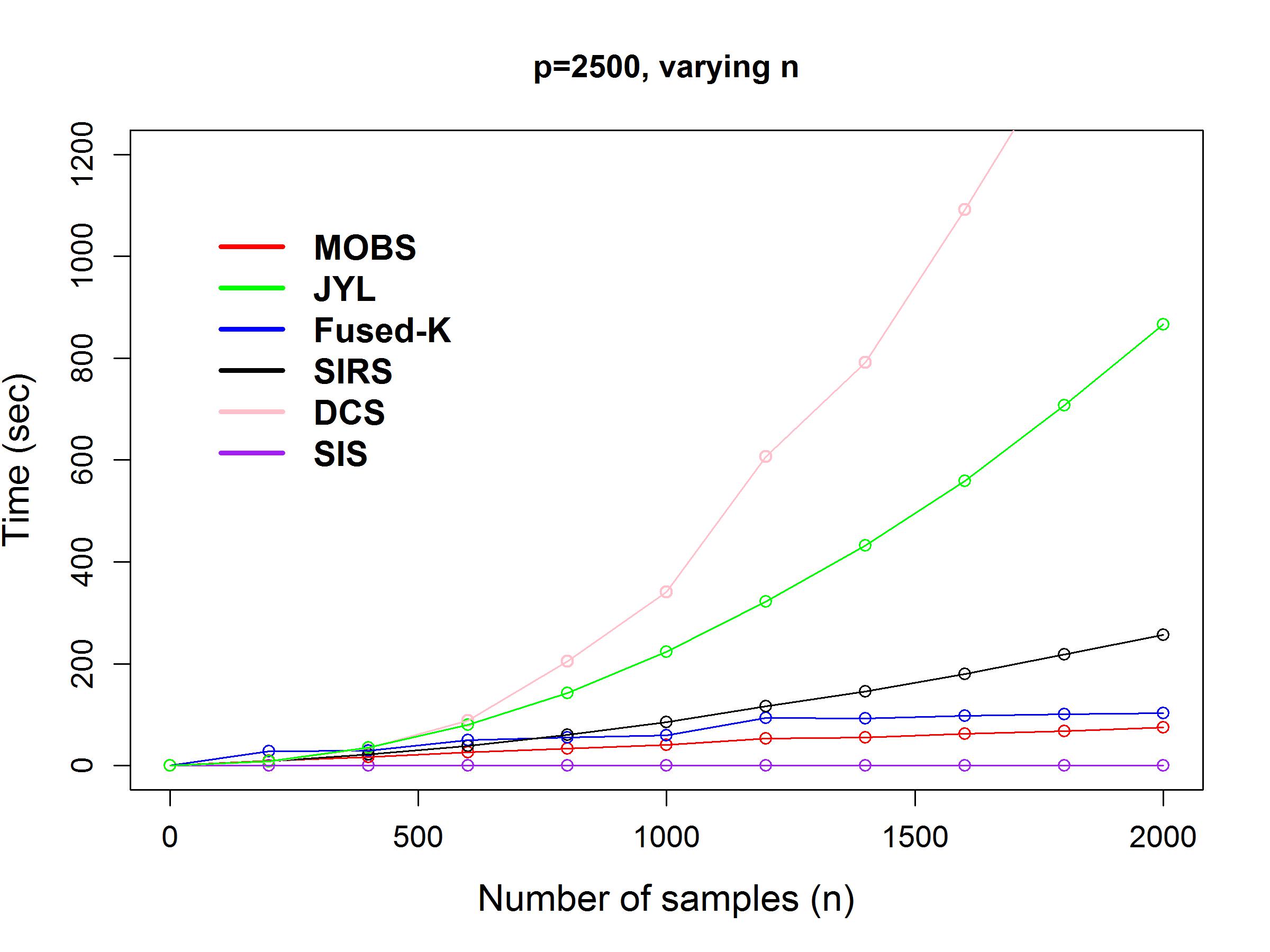}
  \caption{Computational speed against number of samples $n$ when $p=2500$ and computational speed against number of predictors $p$ when $n=250$.}
\end{figure}

Figure 2 summarize the results under both methods. As expected, the linear predictor SIS exhibits significantly faster performance as a simple parametric model though MOBS remains competitive against the other nonparametric and model-free screening methods. All six screening methods are simple to parallelize and scale linearly with the number of predictors $p$. However, in regards to the number of samples $n$, MOBS scales linearly unlike slower competitors such as JYL and DCS.

\section{Applications}
\subsection{Data}
\begin{figure}[h]
\centering
\begin{minipage}{.9\textwidth}
  \includegraphics[width=5.4in]{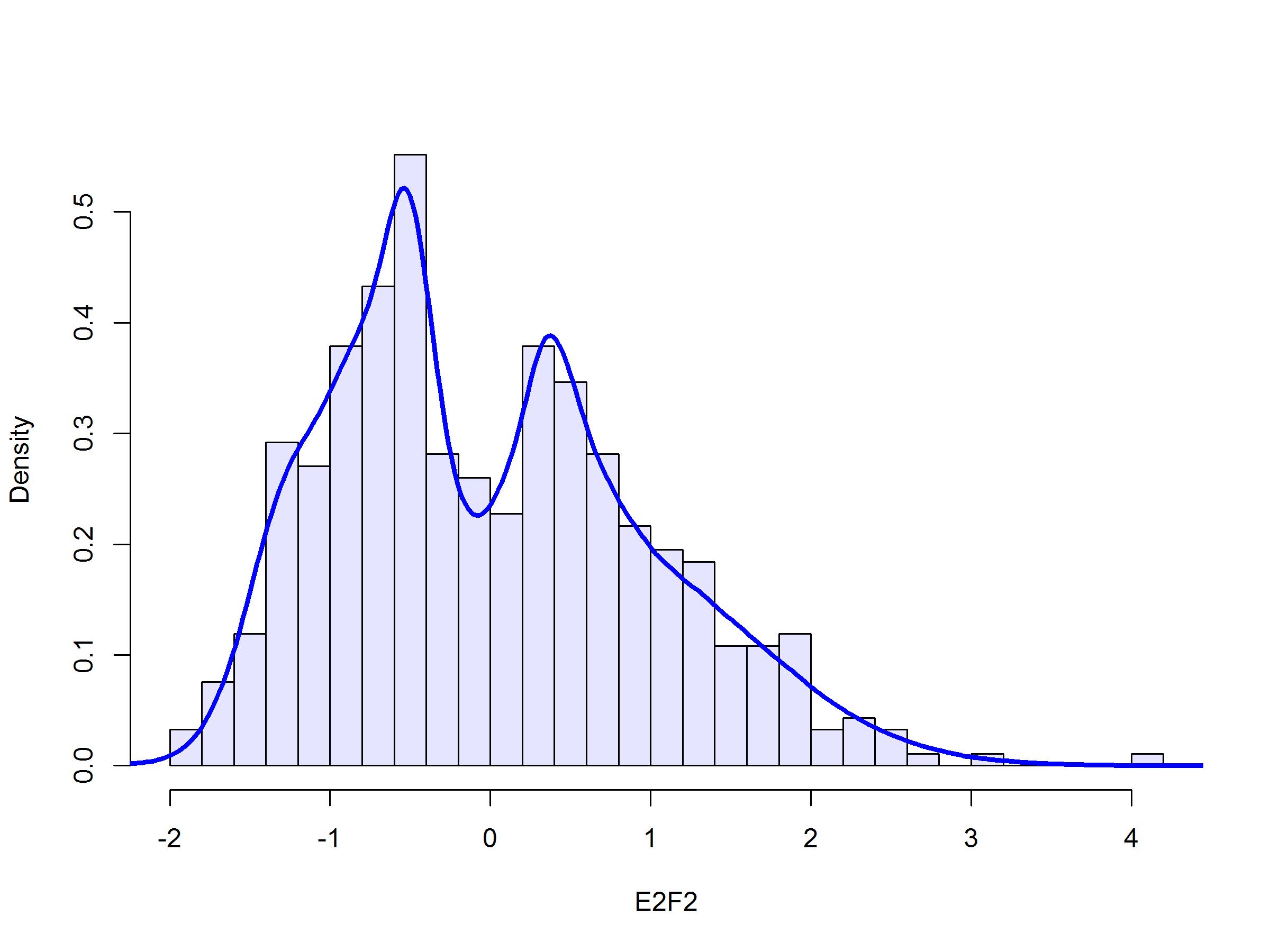}
    \caption{Histogram of $E2F2$ (Ensemble ID: ENSG00000007968) with the estimated density from sampling the baseline model with $k=5$ clusters.}
\end{minipage}%
\end{figure}

In this section, we illustrate our approach by analyzing the GEUVADIS cis-eQTL dataset (\citealt{Lappalainen}), publicly available at \emph{http://www.ebi.ac.uk/Tools/geuvadis-das/}. The data set consists of messenger RNA and microRNA on lymphoblastoid cell line (LCL) samples from 462 individuals provided by the 1000 Genomes Project along with roughly 38 million SNPs. These individuals are taken from 5 different populations: the Yoruba (YRI), CEPH (CEU), Toscani (TSI), British (GBR) and Finns (FIN). We focus on the gene $E2F2$ (Ensemble ID: ENSG00000007968) as our response. $E2F2$ plays a key role in the control of the cell cycle (\citealt{Attwooll}). Traditional eQTL analysis is often limited by the daunting size of the dataset, so our objective is to use screening to select a subset of SNPs that are associated with the gene.

\subsection{Results}
We first ignore the SNPs. We standardize the response $E2F2$ and then run the baseline model under the univariate continuous setting using 7000 iterations with 6500 burn-in and default hyperparameters described in the previous sections with $k=5$ clusters. Figure 3 contains a histogram of the standardized $E2F2$ as well as the estimated marginal density from the Gibbs sampler in the first stage.

Next, we run the second stage and estimate $\mbox{Pr}(H_{0j}| y,x_j,\Lambda,\Phi)$ by averaging $\mbox{Pr}(H_{0j}| y,x_j,\Lambda,\Phi)$ over the baseline $\Phi$ samples, after first removing the SNPs that contain missing data or have $x_j$ values being all 0, all 1 or all 2. The resulting posterior probabilities are displayed in Figure 4. Note that the majority of posterior probabilities are concentrated near 1, with roughly $90\%$ having a posterior probability of 0.95 or greater. On the other hand, about $0.4\%$ of posterior probabilities have values 0.05 or less, with the smallest being $1.864872 \cdot 10^{-8}$. Figure 5 shows the estimated densities at four selected SNPs where red represents samples with $x_j=0$, green with $x_j=1$ and blue with $x_j=2$. The SNPs are chosen with their $\mbox{Pr}(H_{0j}|y,x_j,\Lambda)$ at four different values of under-0.0001, 0.36, 0.61 and over-0.99. Overall, the results suggest that MOBS performs well. SNPs with $\mbox{Pr}(H_{0j}|y,x_j,\Lambda)$ close to 1 have all three estimated distributions nearly identical consistent with the raw data histograms. On the other hand, SNPs with $\mbox{Pr}(H_{0j}|y,x_j,\Lambda)$ close to 0 have three highly varying estimated densities, which are again consistent with smoothed versions of the data histograms.

\begin{figure}[h]
\centering
\begin{minipage}{.9\textwidth}
  \includegraphics[width=5.4in]{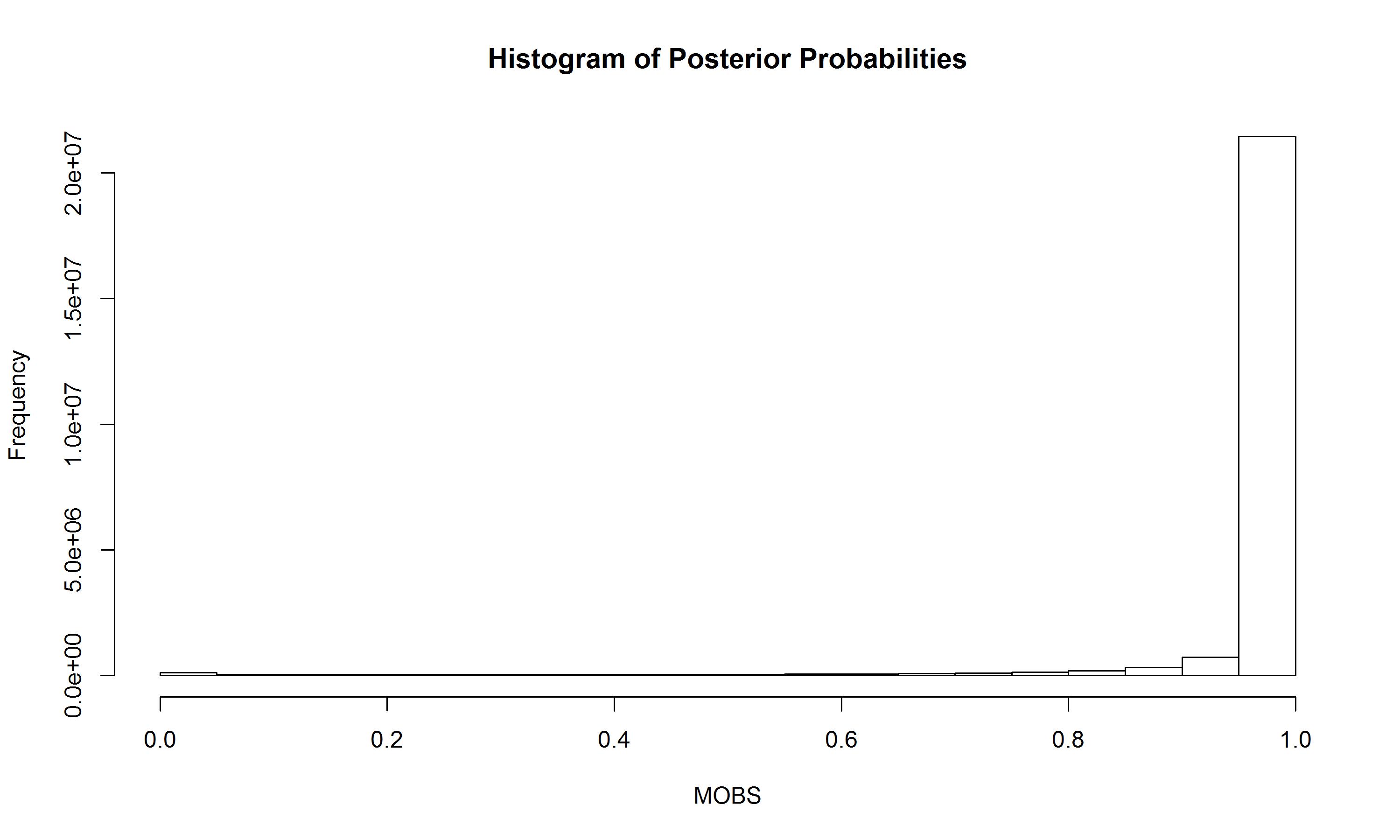}
    \caption{Histogram of $\mbox{Pr}(H_{0j}|X,y)$, measuring the association between $E2F2$ and the SNPs.}
\end{minipage}%
\end{figure}

\begin{figure}[h]
\centering
\begin{minipage}{.9\textwidth}
  \includegraphics[width=6in]{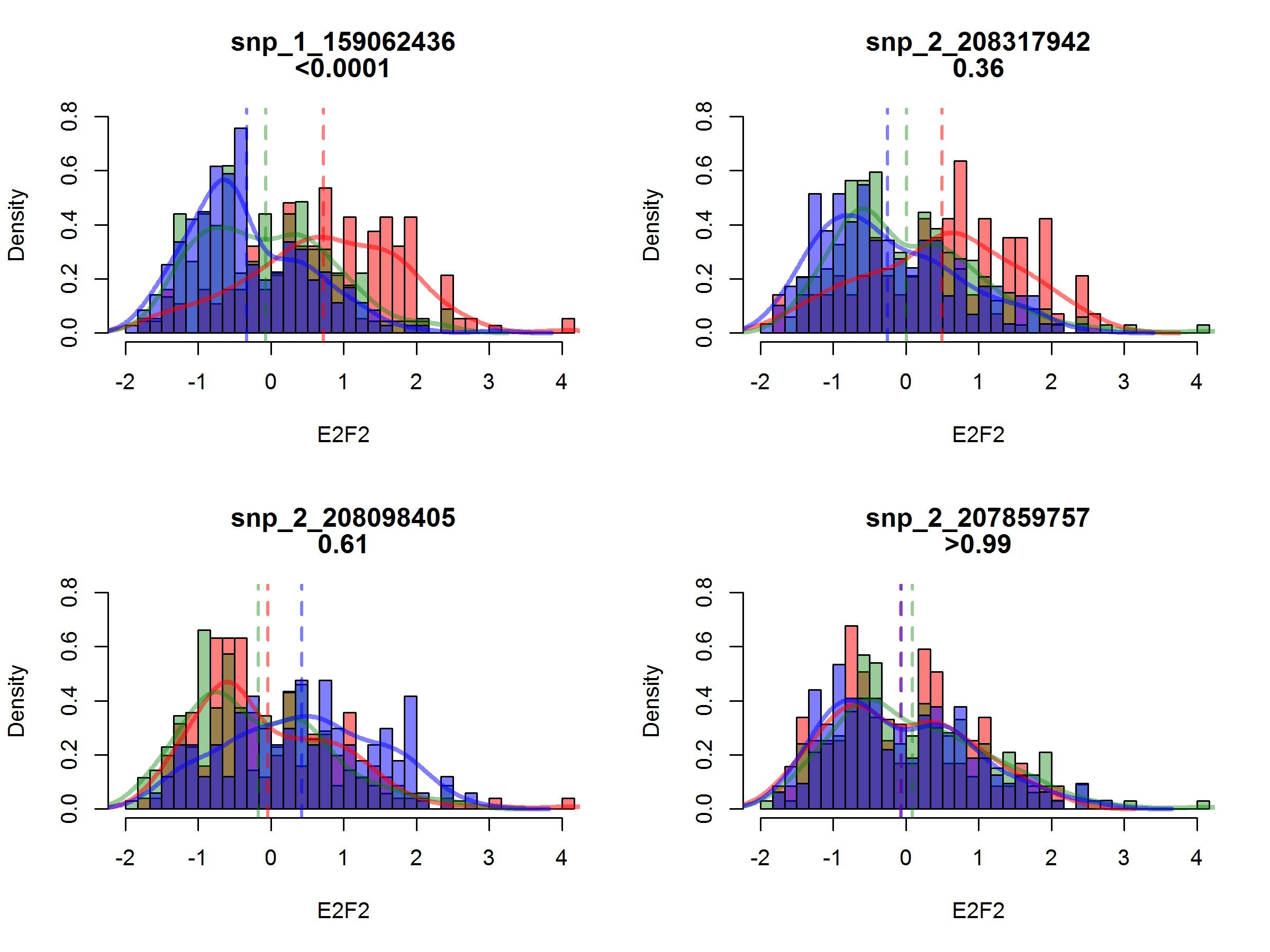}
    \caption{The estimated densities for $E2F2$ for $x_j=0$ (red), $x_j=1$ (green) and $x_j=2$ (blue) at four selected SNPs with varying values of $\mbox{Pr}(H_{0j}|x_j,y,\Lambda)$ at under 0.0001, 0.36, 0.61 and over 0.99. Histograms are also displayed in red, green and blue for each of the three groups.}
\end{minipage}%
\end{figure}

Traditional screening methods generally focus on testing the differences in mean. However, such methods miss more complex shifts in variance or density shape. MOBS is sensitive to not only mean shifts but also shifts in the shape and variance. To judge how much of the SNP selection is due to differences in means, we run the six methods MOBS, DCS, SIS, SIRS, the fused Kolmogorov filter and the JYL Bayesian nonparametric test and then select the 50 most significant predictors across all six models. For DCS, SIS, SIRS, the fused Kolmogorov filter, we convert each $x_j$ into $x^{(0)}_j$ and $x^{(1)}_j$ where $x^{(0)}_j$ and $x^{(1)}_j$ are 0/1 indicators for $x_j=0$ and $x_j=1$ respectively. We then run the models on both $x^{(0)}_j$ and $x^{(1)}_j$ and take the overall test statistic $x_j$ as the maximum of the two. Across all models, we take the sum of the absolute value of the mean for each of the three groups $x_j=0, 1$ and $2$ and display their values in Table 1. The results show that the SNPs selected by DCS, SIS, SIRS and JYL have higher values of total absolute mean distance, which imply that they are predominantly identifying the shifts in the mean. On the other hand, MOBS and FUSEDK have significantly lower values and hence are focusing not just on the mean but on more complex differences.

\begin{table}[htbp]
\centering
\begin{tabular}{| c | c  c  c  c c  c  |}
\hline & MOBS & DCS & SIS & SIRS & FUSEDK & JYL \\\hline
Total Absolute Mean Distance & 1.12 & 1.88 & 1.91 & 1.86 & 1.14 & 1.64 \\\hline
\end{tabular}
\label{tab:atablee}
\caption{The sum of the absolute value of the mean of each cluster for the six models.}
\end{table}

Finally, we assess whether the SNPs identified by MOBS have good predictive performance compared with those identified by DCS, SIS, SIRS, FUSEDK and JYL. We split the data with 370 samples for training and 92 for testing and run 5-fold cross validation. For each screening method, we select the 50 most significant SNPs. In order to examine the predictive capabilities of each method, we fit LASSO after screening. Therefore, the resulting algorithms are MOBS-L, DCS-L, SIS-L, SIRS-L, FUSEDK-L and the JYL-L and their average measure squared errors (MSE) are given in Table 2. While the nonparametric MOBS-L, JYL-L and FUSEDK-L perform better than the SIS-L, DCS-L and SIRS-L, all six methods suffer from poor predictive performance with relatively high MSE. This is not surprising as it tends to be very difficult to accurately predict gene expression based on SNPs alone. The main focus of the analysis is on identifying a promising set of SNPs for further study and not on prediction. A nice feature of MOBS is that it can pick up SNPs different from those identified by existing screening methods.

\begin{table}[htbp]
\centering
\begin{tabular}{| c | c  c  c  c c  c  |}
\hline & MOBS-L & DCS-L & SIS-L & SIRS-L & FUSEDK-L & JYL-L \\\hline
Average MSE & 0.850 & 0.882 & 0.880 & 0.886 & 0.847 & 0.852 \\\hline
\end{tabular}
\label{tab:atablee}
\caption{Comparison of the predictive performance of the six models using the 50 most significant predictors over $E2F2$.}
\end{table}

\section{Discussion}
In this paper, we introduce MOBS, a Bayesian nonparametric class of screening procedures for arbitrary response $y$ and categorical predictors $x_i = (x_{i1},\ldots,x_{ip})'$. By using a type of modularization, MOBS first runs an MCMC algorithm over a baseline model for the marginal response density $f(y)$ and then uses these baseline samples to rapidly estimate each conditional density $f(y|x_j)$. Through simulations, MOBS performs very competitively in terms of both screening performance and computational cost. An analysis on the cis-eQTL genomics dataset reveals that MOBS is able to capture complex shifts beyond a simple difference in means, while maintaining a competitive predictive performance.

The main focus of the paper is on introducing our novel modular screening technique, and providing an illustration through univariate continuous outcome data with massive-dimensional categorical predictors; the cis-eQTL dataset analyzed in Section 5 involved roughly 38 million SNP predictors, which is orders of magnitude larger than is considered in most articles on "high-dimensional" methods.  The proposed MOBS strategy can be applied much more broadly; for example, it is straightforward to consider outcome variables that are multivariate and complex, as long as a Bayesian mixture model and MCMC algorithm has been developed for data of that type.  This is the case not just for multivariate and mixed scale data but also for object data ranging from shapes and curves to graphs.  It is also possible to adapt the methods to accommodate continuous or mixed categorical and continuous predictors. One approach is define $d$ pre-specified knots for each continuous predictor (e.g., spaced at quantiles of the empirical distribution of $x_j$) and use kernels to interpolate $f(y|x_j)$ between these knots.

\centerline{Acknowledgements}

The authors thank Barbara Engelhardt of Princeton University for generously providing the cis-eQTL dataset used in Section 5. This research was partially funded by grant 1546130 of the US National Science Foundation (NSF) and grant 3130624 of the US Office for Naval Research (ONR).

\bibliographystyle{jasa}
\bibliography{mobs-arxiv}
\newpage

\centerline{Appendix}

\vskip 10 pt
\centerline{\emph{Algorithm for univariate Gaussian base mixture model}}
In this section, we provide algorithmic details to sample from the base univariate Gaussian mixture model described in \textbf{Algorithm 1}. Let $n_h=\sum_{i=1}^n 1_{c_i=h}$ be the number of individuals in component $h$ and $\bar{y}_h=\sum_{i:c_i=h} y_i/n_h$ be the mean of $y$ within component $h$. We use a Gibbs sampler to update parameters using the full conditionals:
\begin{enumerate}
\item Sample the cluster allocations from multinomial conditional distributions with $$\mbox{Pr}(c_i=h|-)=\frac{\omega_hN(y_i|\mu_h, \sigma_h^2)}{\sum_{l=1}^k \omega_l N(y_i|\mu_l, \sigma_l^2)},\,\,\,\, h=1,2,\ldots,k$$
\item Sample the component-specific means and variances from
$$\mu_h, \sigma_h^2|- \sim \mbox{N}(\mu_h|\hat{\mu}_h, \hat{q}_h{\sigma}^2_h)  \mbox{IGa}(\sigma_h^2|\hat{a}_h, \hat{b}_h),\,\,\,\, h=1,2,\ldots,k$$
where $$\hat{q}_h=(q^{-1}+n_h)^{-1}, \,\,\,\,\hat{\mu}_h=\hat{q}_h(q^{-1}\mu_0+n_h\bar{y}_h),\,\,\,\, \hat{a}_h=a+n_h/2,$$
$$\hat{b}_h=b+\frac{1}{2}\left[\sum_{i: c_i=h}(y_i-\bar{y}_h)^2+\left(\frac{n_h}{1+qn_h}\right)(\bar{y}_h-\mu_0)^2\right],$$
\item Sample the vector of weights on the different mixture components
$$\omega_1,\ldots,\omega_k|- \sim \mbox{Dir}\left(\frac{\alpha}{k}+n_1,\ldots,\frac{\alpha}{k}+n_k\right).$$
\end{enumerate}

\centerline{\emph{Proof of Theorem 3.1}}

\begin{lem}
As $n_j^0,n_j^1\rightarrow \infty$, with $\frac{n_j^0}{n_j^1} = \frac{\lambda_0}{1-\lambda_0}$ for some fixed $\lambda_0$, we have
$$\mbox{BF}_{11}(j)\sim  C_\omega(j) n^{-(k-1)}\prod_{h=1}^k \left(\frac{p_{jh0}}{\omega_{h}}\right)^{n_{jh}(0)}\left(\frac{p_{jh1}}{\omega_{h}}\right)^{n_{jh}(1)},$$
where
$$C_\omega(j)=\frac{(2\pi)^{k-1}}{\left(\beta(\tau_\omega\omega)\right)^2}\left(\lambda_0(1-\lambda_0)\right)^{-\frac{k-1}{2}}\prod_{h=1}^k \left(p_{jh0}p_{jh1}\right)^{\tau_\omega\omega_{h}-\frac{1}{2}},$$
and
$$\mbox{BF}_{12}(j) \sim  C_\theta(j) e^{-n/2}n^{-2k}\prod_{h=1}^k \left(\frac{n_{jh}(0)}{2b_{jh}(0)}\right)^{a+\frac{n_{jh}(0)}{2}}\left(\frac{n_{jh}(1)}{2b_{jh}(1)}\right)^{a+\frac{n_{jh}(1)}{2}}(\sigma_h^2)^{\frac{n_{jh}(0)+n_{jh}(1)}{2}} e^{\sum_{i: c_i=h} \frac{(y_i-\mu_h)^2}{2\sigma_h^2}},$$
where $$C_\theta(j)=(2\pi)^k 2^{2k}(\lambda_0(1-\lambda_0))^{-k}\tau_\mu^k \prod_{h=1}^k \frac{{b_h}^{2a_h} }{\Gamma(a_h)^{2} p_{jh0} p_{jh1}}.$$
\vskip 10 pt
\end{lem}
\textit{Proof.}
By Stirling's formula, for $c>0$ and $x\rightarrow \infty$, we have
 \begin{eqnarray*}
 \Gamma(x) &\sim & \sqrt{\frac{2\pi}{ x}} \left(\frac{x}{e}\right)^x,\\
 \Gamma(x+c) &\sim & x^c \Gamma(x).
 \end{eqnarray*}
Thus, $$\Gamma(x+c) \sim  \sqrt{2\pi}\,\,\frac{x^{x+c-\frac{1}{2}}}{e^x}.$$
When $n_j^{0}\rightarrow \infty$, we know
\begin{eqnarray*}
\beta(n_j(0)+\tau_\omega\omega) &=&\frac{\prod_{h=1}^k \Gamma(n_{jh}(0)+\tau_\omega\omega_h)}{\Gamma(n_j^0+\tau_\omega)}\\
&\sim& \frac{\prod_{h=1}^k \sqrt{2\pi} \frac{(n_{jh}(0))^{n_{jh}(0)+\tau_\omega\omega_h-\frac{1}{2}}}{e^{n_{jh}(0)}}}{\sqrt{2\pi} \frac{(n_j^0)^{n_j^{0}+\tau_\omega-\frac{1}{2}}}{e^{n_j^{0}}}}\\
&\sim &\left(\frac{2\pi}{n_j^0}\right)^{\frac{k-1}{2}}\frac{\prod_{h=1}^k (n_{jh}(0))^{n_{jh}(0)+\tau_\omega\omega_{h}-1/2}}{(n_j^0)^{n_j^0+\tau_\omega-k/2}}\\
&=& \left(\frac{2\pi}{n_j^0}\right)^{\frac{k-1}{2}} \prod_{h=1}^k p_{jh0}^{n_{jh}(0)+\tau_\omega\omega_{h}-1/2}\\
&=& \left(\frac{2\pi}{n\lambda_0}\right)^{\frac{k-1}{2}} \prod_{h=1}^k p_{jh0}^{n_{jh}(0)+\tau_\omega\omega_{h}-1/2}
\end{eqnarray*}
Therefore,
\begin{eqnarray*}
\mbox{BF}_{11}(j)& =& \frac{\beta(n_j(0)+\tau_\omega\omega)}{\beta(\tau_\omega \omega)} \frac{\beta(n_j(1)+\tau_\omega\omega)}{\beta(\tau_\omega \omega)}\frac{1}{\omega_1^{n_1},\ldots,\omega_k^{n_k}}\\
&\sim & \left[\left(\frac{2\pi}{n\lambda_0}\right)^{\frac{k-1}{2}} \prod_{h=1}^k p_{jh0}^{n_{jh}(0)+\tau_\omega\omega_{h}-1/2} \right] \left[\left(\frac{2\pi}{n(1-\lambda_0)}\right)^{\frac{k-1}{2}} \prod_{h=1}^k p_{jh1}^{n_{jh}(1)+\tau_\omega\omega_{h}-1/2} \right]\\
&&\frac{1}{(\beta(\tau_\omega\omega))^2\omega_1^{n_1},\ldots,\omega_k^{n_k}}\\
&=& C_\omega(j) n^{-(k-1)}\prod_{h=1}^k \left(\frac{p_{jh0}}{\omega_h}\right)^{n_{jh}(0)}\left(\frac{p_{jh1}}{\omega_h}\right)^{n_{jh}(1)},
\end{eqnarray*}
where $$C_\omega(j)=\frac{(2\pi)^{k-1}}{(\lambda_0(1-\lambda_0))^{\frac{k-1}{2}}(\beta(\tau_\omega\omega))^2}\prod_{h=1}^k (p_{jh0}p_{jh1})^{\tau_\omega\omega_{h}-1/2}.$$

As for $\mbox{BF}_{12}(j)$, we see that
\begin{eqnarray*}
\Gamma(a_h+n_{jh}(0)/2) & \sim & \sqrt{2\pi}\,\,\,\frac{(n_{jh}(0)/2)^{a_h+n_{jh}(0)/2-1/2}}{e^{n_{jh}(0)/2}}.
\end{eqnarray*}
Thus
\begin{eqnarray*}
&&\prod_{h=1}^k \Gamma(a_h+n_{jh}(0)/2)\Gamma(a_h+n_{jh}(1)/2)\\
& \sim & {(2\pi)}^{k} e^{-n/2}\prod_{h=1}^k \left(\frac{n_{jh(0)}}{2}\right)^{a_h+n_{jh}(0)/2-1/2}\left(\frac{n_{jh}(1)}{2}\right)^{a_h+n_{jh}(1)/2-1/2}.
\end{eqnarray*}
Also
$$\frac{\tau_\mu}{\sqrt{(\tau_\mu+n_{jh}(0))(\tau_\mu+n_{jh}(1))}}\sim \frac{\tau_\mu}{\sqrt{n_{jh}(0)n_{jh}(1)}}.$$
Thus
\begin{eqnarray*}
\mbox{BF}_{12}(j)&\sim &\frac{ \prod_{h=1}^k\frac{\Gamma(a_{jh}(0))\Gamma(a_{jh}(1))}{\Gamma(a_h)\Gamma(a_h)}\frac{{b_h}^{a_h}{b_h}^{a_h}}{b_{jh}(0)^{a_{jh}(0)}b_{jh}(1)^{a_{jh}(1)}} \frac{\tau_\mu}{\sqrt{n_{jh}(0)n_{jh}(1)}}}{\prod_{h=1}^k \frac{1}{(\sigma_h^2)^{(n_{jh}(0)+n_{jh}(1))/2}} e^{-\sum_{c_i=h} (y_i-\mu_h)^2/(2\sigma_h^2)}}\\
&\sim & C_\theta(j) e^{-n/2}n^{-2k}\frac{\prod_{h=1}^k \left(\frac{n_{jh}(0)}{2b_{jh}(0)}\right)^{a+n_{jh}(0)/2}\left(\frac{n_{jh}(1)}{2b_{jh}(1)}\right)^{a+n_{jh}(1)/2}}{\prod_{h=1}^k \frac{1}{(\sigma_h^2)^{(n_{jh}(0)+n_{jh}(1))/2}} e^{-\sum_{c_i=h} (y_i-\mu_h)^2/(2\sigma_h^2)}}
\end{eqnarray*}
where $$C_\theta(j)=(2\pi)^k 2^{2k}(\lambda_0(1-\lambda_0))^{-k}\tau_\mu^k \prod_{h=1}^k \frac{{b_h}^{2a_h} }{\Gamma(a_h)^{2} p_{jh0} p_{jh1}}.$$ \hfill\fbox{\phantom{\rule{.3ex}{.3ex}}}
\vskip 10 pt

To prove Theorem 3.1, we first set $\lambda_1=1-\lambda_0$ and fix $j=1,\ldots, p$.

We see that for each $h=1,2,..,k$ and $l=0,1$, for given $\Omega_{jh}(l)$, the integer $n_{jh}(l)$ is the number of $\{y_i\}$ with $x_{ij}=l$  in component $h$. As the event that $y_i|x_{ij}=l$ belongs to component $h$ has probability $\Omega_{jh}(l)$, by Central Limit Theorem, we know
 \begin{equation}\label{eq:Zhl}
 p_{jhl}-\Omega_{jh}(l)=O_p\left(\frac{1}{\sqrt{n_j^l}}\right).
 \end{equation}
We  see that for small $t$,
\begin{equation}\label{eq:logexpansion}
(1+t)\log(1+t)=t+O(t^2).
 \end{equation}
 By (\ref{eq:Zhl}), (\ref{eq:logexpansion}), we can get
 \begin{eqnarray}\label{eq:xlogx}
 \frac{p_{jhl}}{\Omega_{jh}(l)}\log\left[\frac{p_{jhl}}{\Omega_{jh}(l)}\right]=\frac{p_{jhl}}{\Omega_{jh}(l)}-1+O_p\left(\frac{1}{n_j^l}\right)
 \end{eqnarray}
 Similarly, \begin{equation}\label{eq:ymean}
 \bar{y}_{jh}(l)-M_{jh}(l)=O_p\left(\frac{1}{\sqrt{n_{jh}(l)}}\right),
\end{equation}
 \begin{equation}\label{eq:yymean}
\frac{1}{n_{jh}(l)}\sum_{i:c_i=h, x_{ij}=l}\left\{y_i-\bar{y}_{jh}(l)\right\}^2=\Sigma^2_{jh}(l)+O_p\left(\frac{1}{\sqrt{n_{jh}(l)}}\right).
\end{equation}
Since
\begin{eqnarray*}
\sum_{i:c_i=h, x_{ij}=l} \{y_i-\bar{y}_{jh}(l)\}^2&=&\sum_{i:c_i=h, x_{ij}=l}\{y_i-M_{jh}(l)\}^2-n_{jh}(l)\{\bar{y}_{jh}(l)-M_{jh}(l)\}^2.
\end{eqnarray*}
By (\ref{eq:ymean} and (\ref{eq:yymean}),
\begin{eqnarray*}
\frac{2b_{jh}(l)}{n_{jh}(l)}&=&\frac{\sum_{i:c_i=h, x_{ij}=l} \{y_i-\bar{y}_{jh}(l)\}^2}{n_{jh}(l)}+O_p\left(\frac{1}{\sqrt{n_{jh}(l)}}\right)\\
&=&\frac{\sum_{i:c_i=h, x_{ij}=l}\{y_i-M_{jh}(l)\}^2}{n_{jh}(l)}-\{\bar{y}_{jh}(l)-M_{jh}(l)\}^2+O_p\left(\frac{1}{\sqrt{n_{jh}(l)}}\right)\\
&=& \Sigma^2_{jh}(l)+O_p\left(\frac{1}{\sqrt{n_{jh}(l)}}\right),
\end{eqnarray*}
and
\begin{eqnarray}
{\frac{n_{jh}(l)}{2}}\log\left[\frac{2b_{jh}(l)}{n_{jh}(l)\Sigma^2_{jh}(l)}\right]
&=& \frac{n_{jh}(l)}{2}\left[\frac{2b_{jh}(l)}{n_{jh}(l)\Sigma^2_{jh}(l)}-1+O_p\left(\frac{1}{n_{jh}}\right)\right]\nonumber\\
 &=&\frac{\sum_{i:c_i=h, x_{ij}=l} \{y_i-M_{jh}(l)\}^2}{2\Sigma^2_{jh}(l)}
-\frac{n_{jh}(l)}{2}+O_p(1)\label{eq:Sigmalogx}
\end{eqnarray}

We now consider the case  $H_{0j}$: $\Omega_{jh}(l)=\Omega_h, M_{jh}(l)=\mu_h, \Sigma_{jh}(l)=\sigma_h$. Since $n_{jh}(l)=p_{jhl}n_j^l,$ we know by (\ref{eq:xlogx}),
\begin{eqnarray*}
\sum_{l=0}^1 \log\left[ \left(\frac{p_{jhl}}{\omega_{h}}\right)^{n_{jh}(l)}\right]
&=&\sum_{l=0}^1 n_j^l p_{jhl}\log\left(\frac{p_{jhl}}{\omega_{h}}\right)\\
&=&\sum_{l=0}^1 n_j^l\omega_h\left\{\frac{p_{jhl}}{\omega_h}-1+O_p\left(\frac{1}{n_j^l}\right)\right\}\\
&=& n_{jh}(0)+n_{jh}(1)-n\omega_h+O_p(1).
\end{eqnarray*}
It can be easily verified that
$$\sum_{h=1}^k\sum_{l=0}^1\left[n_{jh}(0)+n_{jh}(1)-n\omega_h\right]=0.$$
Hence
\begin{equation}\label{eq:sumxlogx}
\sum_{h=1}^k\sum_{l=0}^1 \log\left[ \left(\frac{p_{jhl}}{\omega_{h}}\right)^{n_{jh}(l)}\right]=O_p(1).
\end{equation}

This implies
\begin{eqnarray*}
\log(\mbox{BF}_{11}(j))&\sim & -(k-1)\log(n)+\sum_{h=1}^k\sum_{l=0}^1 \log\left[ \left(\frac{p_{jhl}}{\omega_{h}}\right)^{n_{jh}(l)}\right]\\
&= &-(k-1)\log(n)+O_p(1).
\end{eqnarray*}
Thus $\mbox{BF}_{11}(j)\rightarrow 0$ with rate $n^{-(k-1)}$.

On the other hand, by (\ref{eq:Sigmalogx}),
\begin{eqnarray*}
\log[\mbox{BF}_{12}(j)]& =& -\frac{n}{2}-2k\log(n) +\sum_{h=1}^k \sum_{l=0}^1\log\left[\left\{\frac{n_{jh}(l)}{2b_{jh}(l)}\right\}^{\frac{n_{jh}(l)}{2}}(\sigma^2_{h})^{\frac{n_{jh}(l)}{2}}\right]\\
& & + \sum_{h=1}^k\sum_{i:c_i=h} \frac{(y_i-\mu_h)^2}{2\sigma^2_h}+O_p(1)\\
&=&-\frac{n}{2}-2k\log(n) +\sum_{h=1}^k \sum_{l=0}^1\left\{-\frac{\sum_{i:c_i=h, x_{ij}=l} \{y_i-\mu_h\}^2}{2\sigma_\mu^2}
+\frac{n_{jh}(l)}{2}\right\}\\
&& +\sum_{h=1}^k\sum_{i:c_i=h} \frac{(y_i-\mu_h)^2}{2\sigma^2_h}+O_p(1)\\
&=&-2k\log(n)  +O_p(1).
\end{eqnarray*}

Thus $\mbox{BF}_{11}(j)\rightarrow 0$ with rate $n^{-2k}$.
Combining $\mbox{BF}_{11}$ and $\mbox{BF}_{12}$, then $\mbox{Pr}(H_{1j}|y, \Phi)\rightarrow 0$ with rate  $n^{-(k-1)}$ under $H_{0j}$.

Now consider the case $H_{1j}$.
We know by (\ref{eq:sumxlogx})
\begin{eqnarray*}
\log(\mbox{BF}_{11}(j))&\sim & -(k-1)\log(n)+\sum_{h=1}^k \sum_{l=0}^1 n_{jh}(l)\log\left(\frac{p_{jhl}}{\omega_h}\right)\\
&=&-(k-1)\log(n)+\sum_{h=1}^k\sum_{l=0}^1 n_{jh}(l)\left\{\log\frac{\Omega_{jh}(l)}{\omega_h}+\log\frac{p_{jhl}}{\Omega_{jh}(l)}\right\}\\
&=&-(k-1)\log(n)+n A+O_p(1),
\end{eqnarray*}
where $$A=\sum_{h=1}^k \left[ \lambda_0\Omega_{jh}(0)\log\left(\frac{\Omega_{jh}(0)}{ \omega_h}\right)+\lambda_1\Omega_{jh}(1)\log\left(\frac{\Omega_{jh}(1))}{ \omega_h}\right)\right].$$
The sum $A$ can be seen as Kullback-Leibler divergence between distributions $P=\{\lambda_0\Omega_{jh}(0), \lambda_1 \Omega_{jh}(1)\}_{h=1}^k$ and $Q=\{\lambda_0\omega_h, \lambda_1\omega_h\}_{h=1}^k$. Under the condition $\Omega_{jh}(0)\neq \Omega_{jh}(1)$ for some $h$, the Kullback-Leibler divergence $A>0$. Therefore $\mbox{BF}_{11}(j)$ diverges to $\infty$ with rate $e^{nA} n^{-(k-1)}$.

Next for $\log(\mbox{BF}_{12}(j))$, we know by (\ref{eq:Sigmalogx}) and (\ref{eq:sumxlogx}) that
\begin{eqnarray*}
\log(\mbox{BF}_{12}(j))
&=&  -\frac{n}{2}-2k\log(n) +\sum_{h=1}^k \sum_{l=0}^1\log\left[\left\{\frac{n_{jh}(l)}{2b_{jh}(l)}\right\}^{\frac{n_{jh}(l)}{2}}(\Sigma^2_{jh}(l))^{\frac{n_{jh}(l)}{2}}\right]\\
& & + \sum_{h=1}^k \sum_{l=0}^1 \frac{n_{jh}(l)}{2}\log\left[\frac{\sigma_h^2}{\Sigma^2_{jh}(l)}\right]+ \sum_{h=1}^k\sum_{i:c_i=h} \frac{(y_i-\mu_h)^2}{2\sigma^2_h}+O_p(1)\\
&=&  -\frac{n}{2}-2k\log(n) + \sum_{h=1}^k \sum_{l=0}^1 \frac{n_{jh}(l)}{2}\log\left[\frac{\sigma_h^2}{\Sigma^2_{jh}(l)}\right]\\
&&+ \sum_{h=1}^k\sum_{i:c_i=h} \frac{(y_i-\mu_h)^2}{2\sigma^2_h}+O_p(1).
\end{eqnarray*}

We decompose the sum
\begin{eqnarray*}
 &&\sum_{h=1}^k\sum_{i:c_i=h} \frac{(y_i-\mu_h)^2}{\sigma_h^2}\\
 & = & \sum_{h=1}^k\sum_{l=0}^1 \sum_{i:c_i=h,x_{ij}=l} \left[\frac{\{y_i-M_{jh}(l)\}^2+2\{y_i-M_{jh}(l)\}\{M_{jh}(l)-\mu_h\}+\{M_{jh}(l)-\mu_h\}^2}{\sigma_h^2}\right]\\
&=& \sum_{h=1}^k\sum_{l=0}^1 \left[\frac{n_{jh}(l)\{M_{jh}(l)-\mu_h\}^2+ 2n_{jh}(l)\{\bar{y}_{jh}(l)-M_{jh}(l)\}\{M_{jh}(l)-\mu_h\}}{\sigma_h^2}\right]\\
&&+\sum_{h=1}^k\sum_{l=0}^1 \left[\frac{\Sigma^2_{jh}(l)}{\sigma_h^2} \sum_{i:c_i=h,x_{ij}=l} \frac{\{y_i-M_{jh}(l)\}^2}{\Sigma_{jh}^2(l)}\right]\\
&=& \sum_{h=1}^k\sum_{l=0}^1 \left[\frac{n\lambda_lp_{jhl}\{M_{jh}(l)-\mu_h\}^2}{\sigma_h^2}+ O_p(\sqrt{n})+\frac{\Sigma^2_{jh}(l)}{\sigma_h^2}\left\{n\lambda_lp_{jhl}+O_p(\sqrt{n})\right\}\right]\\
&=& n\sum_{h=1}^k\sum_{l=0}^1 \frac{\lambda_l\Omega_{jh}(l)\{M_{jh}(l)-\mu_h\}^2}{\sigma_h^2}+n\sum_{h=1}^k\sum_{l=0}^1\frac{\Sigma^2_{jh}(l)}{\sigma_h^2}\lambda_l\Omega_{jh}(l)+O_p(\sqrt{n}).
\end{eqnarray*}
Therefore,
\begin{eqnarray*}
\log(\mbox{BF}_{12})&\sim & -2k\log(n) -\frac{n}{2}+ n\sum_{h=1}^k \sum_{l=0}^1 \lambda_l\frac{\Omega_{jh}(l)}{2}\log\left[\frac{\sigma_h^2}{\Sigma^2_{jh}(l)}\right]\\
& &  +  \frac{n}{2}\sum_{h=1}^k\sum_{l=0}^1 \frac{\lambda_l\Omega_{jh}(l)\{M_{jh}(l)-\mu_h\}^2}{\sigma_h^2}+\frac{n}{2}\sum_{h=1}^k\sum_{l=0}^1\frac{\Sigma^2_{jh}(l)}{\sigma_h^2}\lambda_l\Omega_{jh}(l)+O_p(\sqrt{n})\\
& = & nB+O_p(\sqrt{n}),
\end{eqnarray*}
where $$B=\frac{1}{2}\sum_{h=1}^k\sum_{l=0}^1 \frac{\lambda_l\Omega_{jh}(l)\{M_{jh}(l)-\mu_h\}^2}{\sigma_h^2}+\frac{1}{2}\sum_{h=1}^k\sum_{l=0}^1\lambda_l\Omega_{jh}(l)\left[\frac{\Sigma^2_{jh}(l)}{\sigma_h^2}-1+\log\left\{\frac{\sigma_h^2}{\Sigma^2_{jh}(l)}\right\}\right].$$
We see for $t>0$, $$t-1-\log t\geq 0,$$
as $t-1-\log t$  has minimum value 0 at $t=1$. This means that
$$\frac{\Sigma^2_{jh}(l)}{\sigma_h^2}-1+\log\left[\frac{\sigma_h^2}{\Sigma^2_{jh}(l)}\right]\geq 0,$$
with equality holding if and only if $\Sigma^2_{jh}(l)=\sigma_h^2.$
Therefore,  $B\geq 0$ and $B=0$ if and only if $M_{jh}(0)=M_{jh}(1)=\mu_h, \Sigma^2_{jh}(0)=\Sigma^2_{jh}(1)=\sigma_h^2$ for all $h=1,2,...k$. This implies under the condition $\theta_j(0)\neq \theta_j(1)$, $\log\mbox{BF}_{12}(j)\rightarrow \infty$ with rate of $Bn$.
 Under $H_1$, then either $\omega_j(0)\neq \omega_j(1)$ or $\theta_j(0)\neq \theta_j(1)$. In either case, at least one of $\log\mbox{BF}_{11}(j)$ or $\log\mbox{BF}_{12}(j)$ goes to infinity with rate $An$ or $Bn$ and hence $\mbox{Pr}(H_{0j}|y,x_j,\Lambda,\Phi)\rightarrow 0$ with rate of $e^{-n \max(A, B)}$. \hfill\fbox{\phantom{\rule{.3ex}{.3ex}}}
\vskip 10 pt
\centerline{\emph{Proof of Lemma 3.2}}

\begin{lem}. Let $y_1, \ldots , y_n$ be independent with density $f_0$. Assume $f^*= \mbox{argmin}_{f \in \mathbb{F}} KL(f_0||f)$ exists and $P_n(B(\epsilon,f^{*}|f_0)) > 0$ for all $\epsilon > 0$. Then, for any fixed $\epsilon > 0$,

$$P_n(f \in \mathbb{F}: d(f,f^{*})\geq \epsilon|y) \rightarrow 0.$$
\end{lem}
\textit{Proof.}
The proof follows from Theorem 2 from \cite{Lock}. Note that the positive neighborhood condition can be satisfied since $\omega$ has a Dirichlet prior, which induces positive support over $\mathbb{F}$. \hfill\fbox{\phantom{\rule{.3ex}{.3ex}}}

The uniqueness assumption $d(\sum \omega_k f_k(\cdot|\theta_k), f^{*}) = 0$ implies $f=f^{*}$ and $(\omega,\theta)=(\omega^{*},\theta^{*})$. Because $d(\sum \omega_k f_k(\cdot|\theta_k), f^{*})$ is continuous with respect to $\omega, \theta$, there exists $\delta > 0$ such that $d(\sum \omega_k f_k(\cdot|\theta_k), f^*) < \delta$ implies $||(\omega,\theta)-(\omega^{*},\theta^{*})|| < \epsilon$. Hence by the previous lemma:

\begin{eqnarray}
 \nonumber 
  P_n((\omega,\theta) \in (\mathbb{S}^{k-1},\mathbb{R}^{k}): ||(\omega,\theta)-(\omega^{*},\theta^{*})|| \geq \epsilon |y) &\leq&  P_n(f \in \mathbb{F}: d(f,f^{*})\geq \epsilon|y) \\
   &\rightarrow& 0
\end{eqnarray}.   \hfill\fbox{\phantom{\rule{.3ex}{.3ex}}}

\vskip 10 pt
\centerline{\emph{Proof of Theorem 3.3}}
Since $\mbox{Pr}(H_{0j}|y,x_j,\omega,\theta,\Lambda)$ is a continuous function of $\omega$ and $\theta$, and $(\omega, \theta)\rightarrow (\omega^*,\theta^*)$ by the previous theorem, we know that $\mbox{Pr}(H_{0j}|y,x_j,\omega,\theta,\Lambda)-\mbox{Pr}(H_{0j}|y,x_j,\omega^*,\theta^*,\Lambda)\rightarrow 0$ when $n_j^0,n_j^1\rightarrow \infty$.

Under $H_{0j}$, $f^{(0)}=f^{(1)}$, so it follows that $f^{*(0)}=f^{*(1)}=f^*$, and $\omega^{*(0)}=\omega^{*(1)}=\omega^*$, $\theta^{*(0)}=\theta^{*(1)}=\theta^*$. By Theorem 3.1, $\mbox{Pr}(H_{0j}|y,x_j,\omega^*,\theta^*,\Lambda)\rightarrow 1$, thus $\mbox{Pr}(H_{0j}|y,x_j,\omega,\theta,\Lambda)\rightarrow 1$ and hence $\mbox{Pr}(H_{0j}|y,x_j,\Lambda)\rightarrow 1$.

On the other hand, if $f^{*(0)}\neq f^{*(1)}$ under $H_{1j}$, then by Theorem 3.1, $\mbox{Pr}(H_{1j}|y,x_j,\omega^*,\theta^*,\Lambda)\rightarrow 0$, and thus $\mbox{Pr}(H_{0j}|y,x_j,\omega,\theta,\Lambda)\rightarrow 0$ and $\mbox{Pr}(H_{0j}|y,x_j,\Lambda)\rightarrow 0$. \hfill\fbox{\phantom{\rule{.3ex}{.3ex}}}

\vskip 10 pt
\centerline{\emph{Proof of Theorem 3.4}}
By (C1), all jointly important predictors are marginally important which implies that the alternative hypothesis $H_{1j}$ is true for all $j \in D$. Hence by (C2) and Theorem 3.3, $\mbox{Pr}(H_{0j}|y,x_j,\Lambda) = 0$. Thus asymptotically as $n,p \rightarrow \infty$, $P(D \subset \hat{D}) \rightarrow 1$. \hfill\fbox{\phantom{\rule{.3ex}{.3ex}}}

\end{document}